\documentclass[runningheads]{llncs}
\usepackage{caption}
\usepackage{amssymb}
\usepackage{amsmath,bm,mathtools}
\usepackage[pdftex]{graphicx}
\usepackage{url}
\usepackage{subfig}
\usepackage{multirow}
\usepackage{enumitem}
\usepackage{flushend}
\usepackage{xcolor}
\usepackage{bbm}
\usepackage{listings}

\usepackage{algorithm}
\usepackage{algorithmic}

\usepackage{accents}

\usepackage{wrapfig}
\usepackage{adjustbox}
\usepackage{threeparttable}

\newcommand{\reals}{\mathbb{R}}

\allowdisplaybreaks
\newcommand{\RN}[1]{%
  \textup{\uppercase\expandafter{\romannumeral#1}}%
}

\newcommand{\huang}[1]{{{\color{blue} \textbf{(Huang: #1)}}}}
\newcommand{\li}[1]{{{\color{magenta} \textbf{(Li: #1)}}}}

\newcommand{\chen}[1]{{{\color{red} \textbf{(Chen: #1)}}}}

\newcommand{\fan}[1]{{{\color{purple} \textbf{(Fan: #1)}}}}

\newcommand{\vx}{\vec{x}}

\newcommand{\vvu}{\vec{u}}
\newcommand{\vv}{\vec{v}}

\newcommand{\vz}{\vec{z}}
\newcommand{\vy}{\vec{y}}

\newcommand{\commentout}[1]{}

\newenvironment{myenumerate}{\begin{list}{$\arabic$}
{\setlength{\topsep}{1mm}
\setlength{\itemsep}{0.25mm}
\setlength{\parsep}{0.25mm}
\setlength{\itemindent}{0mm}
\setlength{\partopsep}{0mm}
\setlength{\labelwidth}{15mm}
\setlength{\leftmargin}{4mm}}}{\end{list}}

\definecolor{codegreen}{rgb}{0,0.6,0}
\definecolor{codegray}{rgb}{0.5,0.5,0.5}
\definecolor{codepurple}{rgb}{0.58,0,0.82}
\definecolor{backcolour}{rgb}{0.95,0.95,0.92}
\lstdefinestyle{mystyle}{
    backgroundcolor=\color{backcolour},   
    commentstyle=\color{codegreen},
    keywordstyle=\color{magenta},
    numberstyle=\tiny\color{codegray},
    stringstyle=\color{codepurple},
    basicstyle=\ttfamily\scriptsize,
    breakatwhitespace=false,         
    breaklines=true,                 
    captionpos=b,                    
    keepspaces=true,                 
    numbers=left,                    
    numbersep=5pt,                  
    showspaces=false,                
    showstringspaces=false,
    showtabs=false,                  
    tabsize=2
}

\newcommand{\relu}{\text{ReLU}}

\usepackage{xcolor}
\definecolor{darkgreen}{rgb}{0,0.5,0}
\definecolor{purple}{rgb}{1,0,1}
\newcommand{\kibitz}[2]{\ifnum\Comments=0\textcolor{#1}{#2}\fi}

\begin{document}
\title{POLAR: A Polynomial Arithmetic\\ Framework for Verifying\\ Neural-Network Controlled Systems}

\author{Chao Huang\inst{1} \and Jiameng Fan\inst{2}\and Zhilu Wang\inst{4} \and Yixuan Wang\inst{4} \and Weichao Zhou\inst{2} \and Jiajun Li\inst{1} \and Xin Chen\inst{3} \and Wenchao Li\inst{2} \and Qi Zhu\inst{4}}

\titlerunning{POLAR}

\institute{University of Liverpool, \email{\{chao.huang2, j.li234\}@liverpool.ac.uk} \and Boston University, \email{\{jmfan, zwc662, wenchao\}@bu.edu} \and University of Dayton, \email{xchen4@udayton.edu} \and Northwestern University, \email{\{yixuanwang2024, zhilu.wang\}@u.northwestern.edu, qzhu@northwestern.edu}}

\maketitle

\begin{abstract}
We present POLAR\footnote{The source code can be found in \url{https://github.com/ChaoHuang2018/POLAR_Tool}.}, a \textbf{pol}ynomial \textbf{ar}ithmetic-based framework 
for efficient bounded-time reachability analysis of neural-network controlled systems (NNCSs). 
Existing approaches that leverage the standard Taylor Model (TM) arithmetic for approximating the neural-network controller cannot deal with non-differentiable activation functions and suffer from rapid explosion of the remainder when propagating the TMs.
POLAR overcomes these shortcomings by
integrating TM arithmetic with \textbf{Bernstein B{\'e}zier Form} and \textbf{symbolic remainder}. 
The former enables TM propagation across non-differentiable activation functions and local refinement of TMs, and the latter reduces error accumulation in the TM remainder for linear mappings in the network. 
Experimental results show that POLAR significantly outperforms the current state-of-the-art tools in terms of both efficiency and tightness of the reachable set overapproximation. 
\end{abstract}

\section{Introduction}

\begin{figure}[htbp]
\centering
\begin{minipage}{0.35\textwidth}
  \centering
     \includegraphics[width=0.7\textwidth]{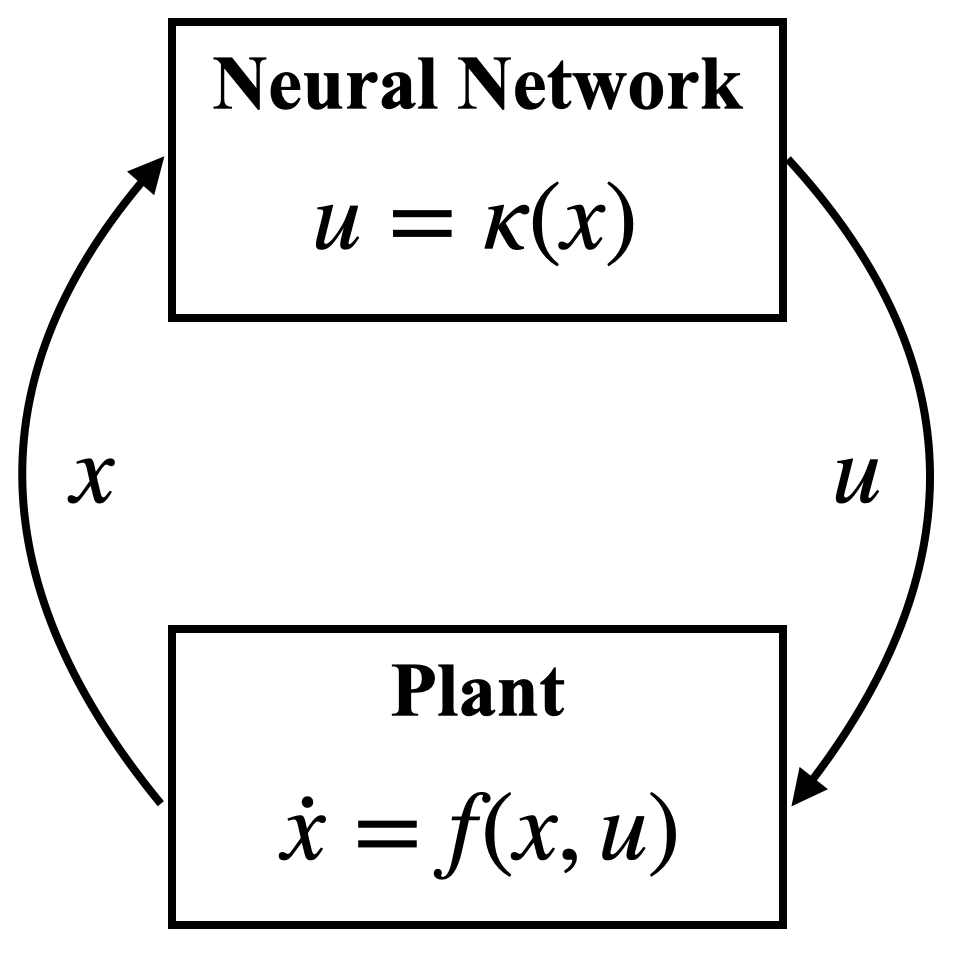}%
	\captionof{figure}{A typical NNCS model.}
	\label{fig:nncs_structure}
\end{minipage}
\hspace{1ex}
\begin{minipage}{0.6\textwidth}
\centering
     \includegraphics[width=0.8\textwidth]{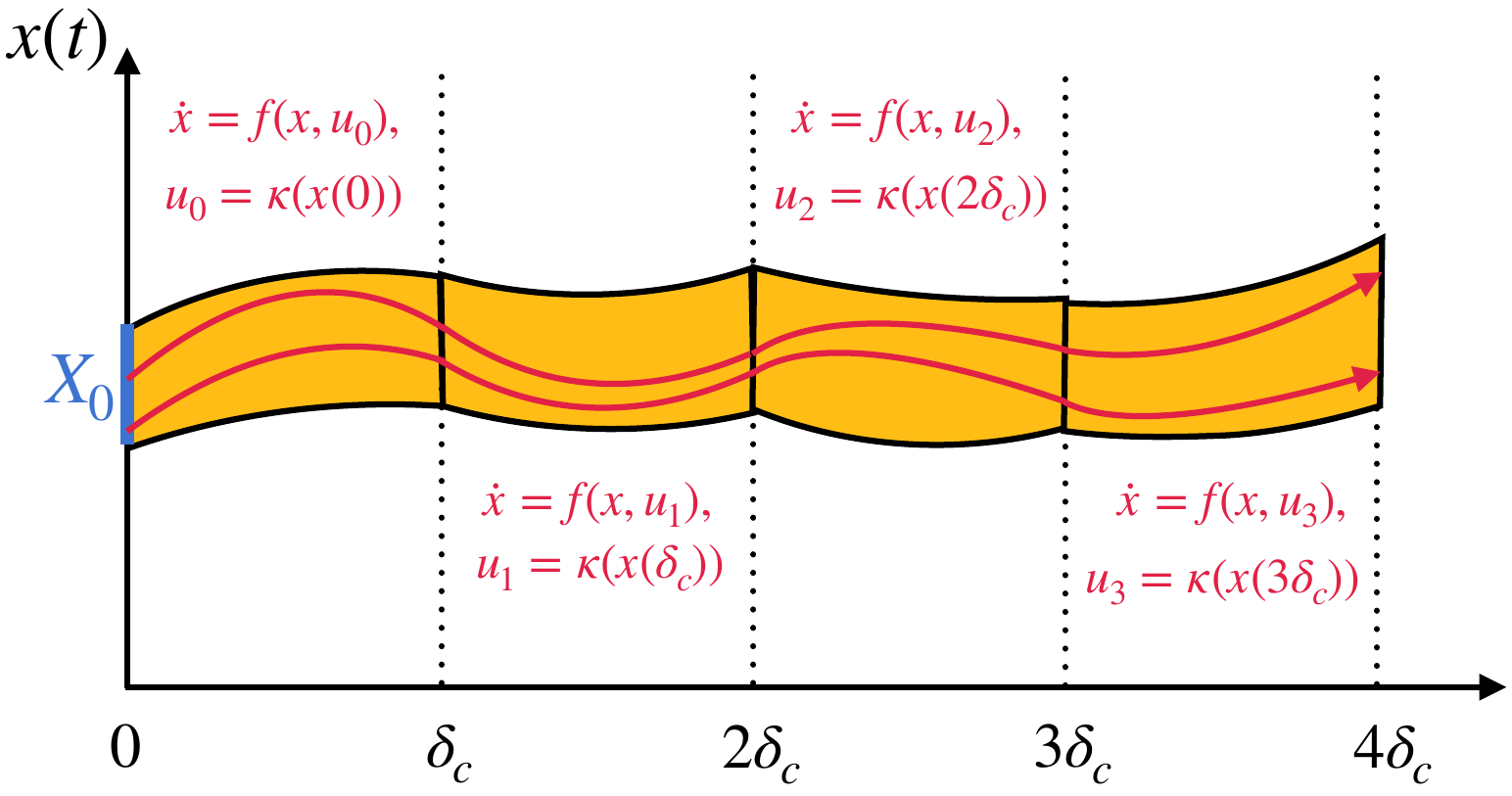}%
	\caption{figure}{Executions over $4$ control steps.}
	\label{fig:flowpipe}
\end{minipage}
\vspace{-0.8cm}
\end{figure}

Neural networks have been increasingly used as the central decision makers in a variety of control tasks~\cite{mnih2015human,pan2018agile,levine2016end}.
However, the use of neural-network controllers also gives rise to new challenges on verifying the correctness of the resulting closed-loop control systems especially in safety-critical settings.
In this paper, we consider the reachability verification problem of neural-network controlled systems (NNCSs). 
The high-level architecture of a simple NNCS is shown in Figure~\ref{fig:nncs_structure} in which the neural network senses the system state, i.e. the value of $\vx$, at discrete time steps, and computes the corresponding control values $\vvu$ for updating the system dynamics which is defined by an ordinary differential equation (ODE) over $\vx$ and $\vvu$.
The \textit{bounded-time reachability analysis problem} of an NNCS is to compute an (overapproximated) reachable set that contains all the trajectories starting from an initial set for a finite number of control steps.
The initial set can represent uncertainties in the starting state of the system or error (e.g. localization error) bounds in estimating the current system state during an execution of the system. 
Figure~\ref{fig:flowpipe} shows an illustration of reachable sets for 4 steps, where the orange region represents the reachable set, and the two red, arrowed curves are two example trajectories starting from two different initial states in the initial set $X_0$ (blue).

Reachability analysis of general NNCSs is notoriously difficult due to nonlinearity in both the neural-network controller and the plant. The difficulty is further exacerbated by the coupling of the controller and the plant over multiple control steps. 
Since exact reachability of general nonlinear systems is undecidable~\cite{Alur+Dill/1994/timed_automata}, current approaches for reachability analysis of nonlinear dynamical systems 
largely focus on computing a tight overapproximation of the reachable sets~\cite{lygeros1999controllers,prajna2004safety,Frehse+/2011/SpaceEx,Chen+/2013/flowstar,Althoff/2015/CORA}.
Verisig~\cite{IvanovWAPL19} leverages properties of the sigmoid activation function and converts an NNCS with these activation functions to an equivalent hybrid system. Thus, existing tools for hybrid system reachability analysis can be directly applied to solve the NNCS reachability problem. However, this approach inherits the efficiency problem of hybrid system reachability analysis and does not scale beyond very small NNCSs. 
Another line of approach is to draw on techniques for computing the output ranges of neural networks~\cite{HuangKWW17,KatzBDJK17,WangPWYJ18,weng2018towards,zhang2018efficient,SinghGPV19} by directly integrating them with reachability analysis tools designed for dynamical systems.
NNV~\cite{TranYLMNXBJ20}, for instance, combines star set analysis on the neural network with zonotope-based analysis of the nonlinear plant dynamics from CORA~\cite{Althoff/2015/CORA}.
However, this type of approach has been shown to be ineffective for NNCS verification due to the lack of consideration on the interaction between the neural-network controller and the plant dynamics~\cite{DuttaCS19,HuangFLC019,ivanov2021veification}.
In particular, since the primary goal of these techniques is to bound the output range of the neural network instead of approximating its input-output function, they \textit{cannot track state dependencies across the closed-loop system and across multiple time steps in reachability analysis}.

More recent advances in NNCS reachability analysis 
are based on the idea of \textit{function overapproximation} of the neural network controller.
A function overapproximation of a neural network $\kappa$ has two components: an approximated function $p$ and an error term $I$ (e.g. an interval) that bounds the approximation error.
Such function overapproximation that produces a \textit{point-wise} approximation of $\kappa$ with an interval error term (typically called a remainder) is also known as a \textit{Taylor model} (TM).
Function-overapproximation approaches
can be broadly categorized into two classes:
\textit{direct end-to-end approximation} such as Sherlock~\cite{DuttaCS19}, ReachNN~\cite{HuangFLC019} and ReachNN*~\cite{FanHCL020}, and \textit{layer-by-layer propagation} such as Verisig 2.0~\cite{IvanovCWAPL21,ivanov2021veification}. The former computes a function overapproximation of the neural network end-to-end by sampling from the input space. The main drawback of this approach is that it does not scale beyond systems with more than a few input dimensions. 
The latter approach tries to exploit the neural network structure and uses \textit{Taylor model arithmetic} to more efficiently obtain a function overapproximation of $\kappa$ by propagating the TMs layer by layer through the network (details in Section~\ref{sec:arithmetic}).
However, due to limitations of basic TM arithmetic, these approaches \textit{cannot handle non-differentiable activation functions and suffer from rapid growth of the remainder} during propagation.
For instance, explosion of the interval remainder would degrade a TM propagation to an interval analysis.

In this paper, we propose a principled \textbf{pol}ynomial \textbf{ar}ithmetic framework (POLAR) that enables precise layer-by-layer propagation of TMs for general feed-forward neural networks.
Basic Taylor model arithmetic cannot handle ReLU that is non-differentiable (cannot produce the polynomial), and also suffers from low approximation precision (large remainder).
POLAR addresses the key challenges of applying basic TM arithmetic through a novel use of \textit{univariate Bernstein approximation} and \textit{symbolic remainders}.
Univariate Bernstein polynomial enables the handling of non-differentiable activation functions and local refinement of Taylor models (details in Section~\ref{sec:main}). 
Symbolic remainders can taper the growth of interval remainders by avoiding the so-called wrapping effect~\cite{Jaulin+/2001/applied_interval_analysis} in linear mappings. The paper has the following novel contributions: (I) A polynomial arithmetic framework using both Taylor and univariate Bernstein approximations for computing NNCS reachable sets to handle general NN controllers; (II) An adaptation of the symbolic remainder method for ODEs to the layer-by-layer propagation for neural networks; (III) A comprehensive experimental evaluation of our approach on challenging case studies that demonstrates significant improvements of POLAR against SOTA.

\section{Preliminaries}\label{sec:pre}





A \emph{Neural-Network Controlled System (NNCS)} is a continuous plant governed by a neural network controller. The plant dynamics is defined by an ODE of the form $\dot{\vx} = f(\vx,\vvu)$ wherein the state variables and control inputs are denoted by the vectors $\vx$ and $\vvu$ respectively. We assume that the function $f$ is at least locally Lipschitz continuous such that its solution w.r.t. an initial state and constant control inputs is unique~\cite{Meiss/2007/Differential}. We denote the input-output mapping of the neural network controller as $\kappa$. The controller is triggered every $\delta_c$ time which is also called the \emph{control stepsize}. A system \emph{execution (trajectory)} is produced in the following way: starting from an initial state $\vx(0)$, the controller senses the system state at the beginning of every control step $t {=} j\delta_c$ for $j{=}0,1,{\dots}$, and updates the control inputs to $\vv_j {=} \kappa(\vx(j\delta_c))$. The system's dynamics in that control step is governed by the ODE $\dot{\vx} {=} f(\vx,\vv_j)$.

Given an initial state set $X_0 \subset \reals^n$, all executions from a state in this set can be formally defined by a \emph{flowmap} function $\varphi_{\mathcal{N}}: X_0 \times \reals_{\geq 0} \rightarrow \reals^n$, such that the system state at any time $t \geq 0$ from any initial state $\vx_0\in X_0$ is $\varphi_{\mathcal{N}}(\vx_0,t)$. 
We call a state $\vx'\in \reals^n$ \emph{reachable} if there exists $\vx_0\in X_0$ and $t \geq 0$ such that $\vx' = \varphi_{\mathcal{N}}(\vx_0,t)$. The \emph{reachability problem} on NNCS is to decide whether a state is reachable in a given NNCS, and it is \emph{undecidable} since NNCS is more expressive than two-counter machines for which the reachability problem is already undecidable~\cite{Alur+Dill/1994/timed_automata}.
Many formal verification problems can be reduced to the reachability problem. 
For example, the safety verification problem can be reduced to checking reachability to an unsafe state. 
In the paper, we focus on computing the reachable set for an NNCS over a bounded number $K$ of control steps. Since flowmap $\varphi_{\mathcal{N}}$ often does not have a closed form due to the nonlinear ODEs, we seek to compute \emph{state-wise overapproximations} for it over multiple time segments, that is, in each control step $[j\delta_c,(j+1)\delta_c]$ for $j=0,\dots,K-1$, the reachable set is overapproximated by a group of flowpipes $\mathcal{F}_1(\vx_0,\tau),\dots,\mathcal{F}_N(\vx_0,\tau)$ over the $N$ uniformly subdivided time segments of the time interval, such that $\mathcal{F}_i(\vx_0,\tau)$ is a \emph{state-wise overapproximation} of $\varphi_{\mathcal{N}}(\vx_0, j\delta_c + (i-1)\delta + \tau)$ for $\tau \in [0,\delta_c/N]$, i.e., $\mathcal{F}_j(\vx_0,\tau)$ contains the exact reachable state from any initial state $\vx_0$ in the $i$-th time segment of the $j$-th control step. Here, $\tau$ is the local time variable which is independent in each flowpipe. A high-level flowpipe construction algorithm is presented as follows, in which $\hat{X}_0 = X_0$ and $\delta = \delta_c/N$ is called the \emph{time step}.

\vspace{1mm}
\begin{algorithmic}[1]
  \FOR{$j=0$ to $K-1$}
   \STATE Computing an overapproximation $\hat{U}_j$ for the control input range $\kappa(\hat{X}_j)$;
   \STATE Computing the flowpipes $\mathcal{F}_1(\vx_0,\tau),\dots,\mathcal{F}_N(\vx_0,\tau)$ for the continuous dynamics $\dot{\vx} = f(\vx,\vvu),\dot{\vvu}=0$ from the initial set $\vx(0) \in \hat{X}_j$, $\vvu(0)\in \hat{U}_j$;
   \STATE $\mathcal{R} \leftarrow \mathcal{R} \cup \{(\mathcal{F}_1(\vx_0,\tau),\dots,\mathcal{F}_N(\vx_0,\tau)\}$;
   \STATE $\hat{X}_{j+1}\ \leftarrow\ \mathcal{F}_N(\vz, \delta)$;
  \ENDFOR
\end{algorithmic}
\vspace{1mm}

Notice that $\vx(0)$ denotes the local initial set for the ODE used in the current control step, that is the system reachable set at the time $j\delta_c$, while the variables $\vx_0$ in a flowpipe are the symbolic representation of an initial state in $X_0$. Intuitively, a flowpipe overapproximates not only the reachable set in a time step, but also the \textit{dependency} from an initial state to its reachable state at a particular time.
For settings where the plant dynamics of an NNCS is given as a difference equation in the form of $\vx_{k+1} = f(\vx_k,\vvu_k)$, we can obtain \textit{discrete} flowpipes which are the reachable set overapproximations at discrete time points by repeatedly computing the state set at the next step using TM arithmetic.

\noindent\textbf{Dependencies on the initial set.}
As we mentioned previously, the reachable state of an NNCS at a time $t>0$ is \textit{uniquely determined} by its initial state if there is no noise or disturbance in the system dynamics or on the state measurements. If we use $X_j$ to denote the exact reachable set $\{\varphi_{\mathcal{N}}(\vx_0, j\delta_c)\,|\,\vx_0\in X_0\}$ from a given initial set $X_0$, then the control input range is defined by the set $U_j = \{\kappa(\vx_j)\,|\,\vx_j = \varphi_{\mathcal{N}}(\vx_0, j\delta_c) \text{ and } \vx_0\in X_0\}$. More intuitively, the set $U_j$ is the image from the initial set $X_0$ under the mapping $\kappa(\varphi_{\mathcal{N}}(\cdot, j\delta_c))$. 
\emph{The main challenge in computing NNCS reachable sets is to control the overapproximation, which requires accurately tracking the dependency of a reachable set on the initial set across multiple control steps.} In this paper, we present a polynomial arithmetic framework for tracking such dependencies using Taylor models.

\noindent\textbf{Taylor model arithmetic.} Taylor models are originally proposed to compute higher-order overapproximations for the ranges of continuous functions (see~\cite{Berz+Makino/1998/Verified}). They can be viewed as a higher-order extension of intervals~\cite{Moore+Others/2009/Interval}, which are sets of real numbers between lower and upper real bounds, e.g., the interval $[a,b]$ wherein $a\leq b$ represents the set of $\{x\,|\,a\leq x\leq b\}$. A \emph{Taylor model (TM)} is a pair $(p,I)$ wherein $p$ is a polynomial of degree $k$ over a finite group of variables $x_1,\dots,x_n$ ranging in an interval domain $D\subset \reals^n$, and $I$ is the remainder interval. The range of a TM is the Minkowski sum of the range of its polynomial and the remainder interval. Thereby we sometimes intuitively denote a TM $(p,I)$ by $p + I$ in the paper.
TMs are closed under  operations such as addition,
multiplication, and integration (see~\cite{Makino+Berz/2003/Taylor}). Given functions
$f,g$ that are overapproximated by TMs $(p_f,I_f)$ and $(p_g,I_g)$,
respectively, a TM for $f+g$ can be computed as $(p_f + p_g, I_f +
I_g)$, and an order $k$ TM for $f\cdot g$ can be computed as $( \, p_f
\cdot p_g - r_k \, , \, I_f \cdot B(p_g) + B(p_f) \cdot I_g +
I_f{\cdot} I_g + B(r_k) \, )$, wherein $B(p)$ denotes an interval
enclosure of the range of $p$, and the \emph{truncated part} $r_k$
consists of the terms in $p_f\cdot p_g$ of degrees $> k$. Similar to reals and intervals, TMs can also be organized as vectors and matrices to overapproximate the functions whose ranges are multidimensional.
Notice that \emph{a TM is a function overapproximation and not just a range overapproximation like intervals or polyhedra.}

\section{Framework of POLAR}
\label{sec:arithmetic}

In this section, 
we describe POLAR's approach for computing a TM for the output range of a neural network (NN) when the input range is defined by a TM. POLAR uses the layer-by-layer propagation strategy, and features the following key novelties:
(a) A method to compute univariate Bernstein Polynomial \textbf{(BP)} overapproximations for activation functions, and selectively uses Taylor or Bernstein polynomials to \emph{limit the overestimation produced when overapproximating the output ranges of individual neurons}.
(b) A technique to symbolically represent the intermediate linear transformations of TM interval remainders during the layer-by-layer propagation. The purpose of using Symbolic Remainders \textbf{(SR)} is to \emph{reduce the accumulation of overestimation in composing a sequence of TMs}.

\subsection{Main Framework} \label{sec:main}

We begin by introducing POLAR's propagation framework that incorporates only (a), and then describe how to extend it by further integrating (b). Although using TMs to represent sets in layer-by-layer propagation is already used in~\cite{IvanovCWAPL21,ivanov2021veification}, the method only computes Taylor approximations for activation functions, and the TM output of one layer is propagated by the existing arithmetic for TM composition to the next layer. Such a method has the following shortcomings: (1) the activation functions have to be differentiable, (2) standard TM composition is often the source of overestimation even preconditioning and shrink wrapping are used. Here, we seek to improve the use of TMs in the above two aspects.

\begin{wrapfigure}{r}{0.41\textwidth}
  \vspace{-0.2cm}
  \begin{center}
    \includegraphics[width=0.4\textwidth]{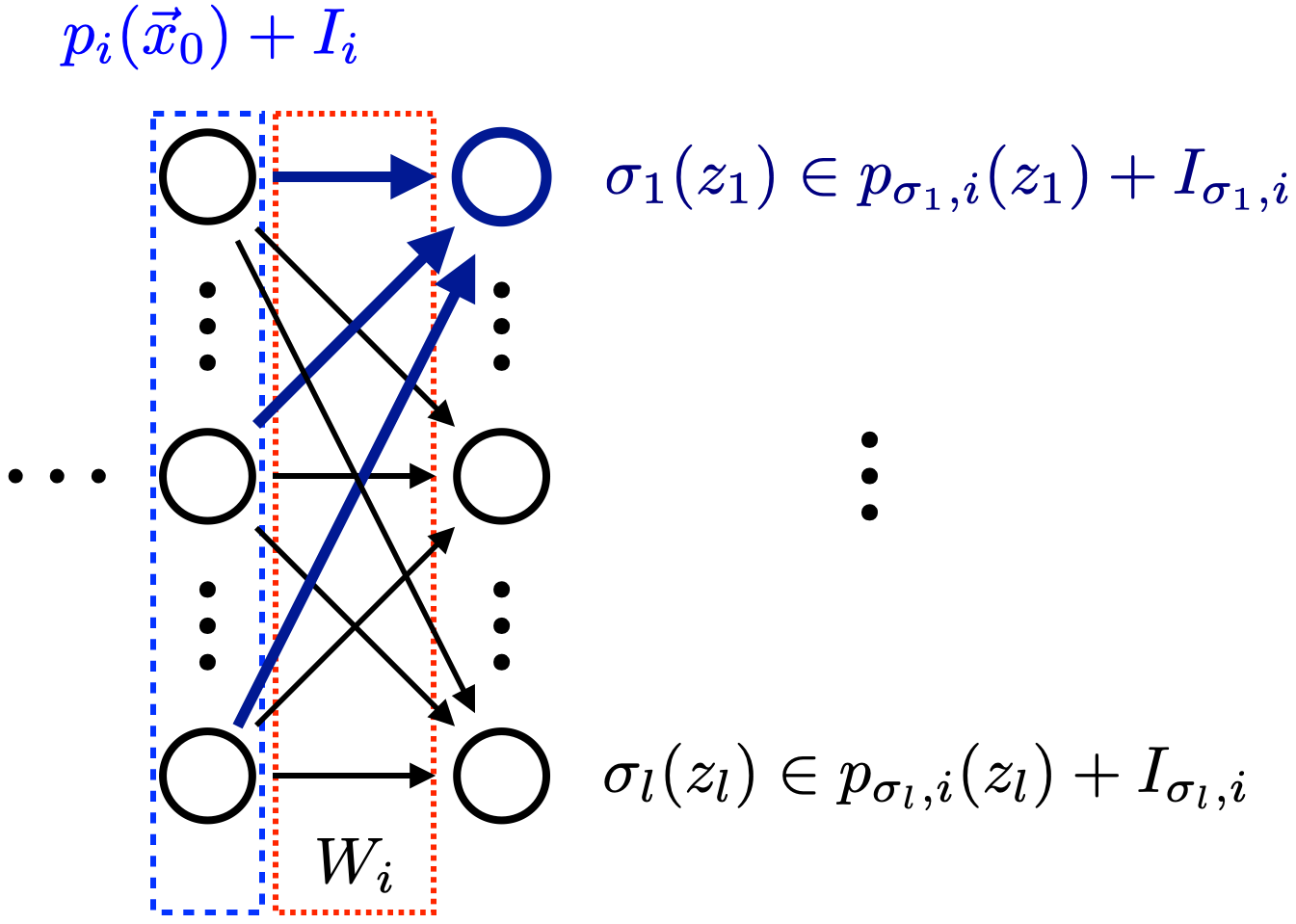}
  \end{center}
  \caption{Single layer propagation}\label{fig:single_layer_composition}
  \vspace{-0.5cm}
\end{wrapfigure}

Before presenting our layer-by-layer propagation method, we describe how a TM output is computed from a given TM input for a single layer. The idea is illustrated in Fig.~\ref{fig:single_layer_composition}. The circles in the right column denote the neurons in the current layer which is the $i$-th layer, and those in the left column denotes the neurons in the previous layer. The weights on the incoming edges to the current layer is organized as a matrix $W_i$, while we use $B_i$ to denote the vector organization of the biases in the current layer. Given that the output range of the neurons in the previous layer is represented as a TM (vector) $(p_i(\vx_0), I_i)$ wherein $\vx_0$ are the variables ranging in the NNCS initial set. Then, the output TM $(p_{i+1}(\vx_0), I_{i+1})$ of the current layer can be obtained as follows. First, we compute the polynomial approximations $p_{\sigma_1,i},\dots,p_{\sigma_l,i}$ for the activation functions $\sigma_1,\dots,\sigma_l$ of the neurons in the current layer. Second, interval remainders $I_{\sigma_1,i},\dots,I_{\sigma_l,i}$ are evaluated for those polynomials to ensure that for each $j=1,\dots,l$, $(p_{\sigma_j,i},I_{\sigma_j,i})$ is a TM of the activation function $\sigma_j$ w.r.t. $z_j$ ranging in the $j$-th dimension of the set $W_i(p_i(\vx_0) + I_i)$. Third, $(p_{i+1}(\vx_0,I_{i+1}))$ is computed as the TM composition $p_{\sigma,i}(W_i(p_i(\vx_0) + I_i) + I_{\sigma,i}$ wherein $p_{\sigma,i}(\vz) = (p_{\sigma_1,i}(z_1),\dots,p_{\sigma_l,i}(z_k))^T$ and $I_{\sigma,i} = (I_{\sigma_1,i},\dots,I_{\sigma_l,i})^T$. Hence, when there are multiple layers, starting from the first layer, the output TM of a layer is treated as the input TM of the next layer, and the final output TM is computed by composing TMs layer-by-layer.

\begin{algorithm} [tbp]
 \caption{Layer-by-layer propagation using polynomial arithmetic and TMs}\label{algo:nn_output}
 \begin{algorithmic}[1]
  \REQUIRE Input TM $(p_1(\vx_0),I_1)$ with $\vx_0\in X_0$, the $M+1$ matrices $W_1,\dots,W_{M+1}$ of the weights on the incoming edges of the hidden and the output layers, the $M+1$ vectors $B_1,\dots,B_{M+1}$ of the neurons' bias in the hidden and the output layers, the $M+1$ activation functions $\sigma_1,\dots,\sigma_{M+1}$ of hidden and output layers.
  \ENSURE a TM $(p_r(\vx_0),I_r)$ that contains the set $\kappa((p_1(\vx_0),I_1))$.
  \STATE $(p_r,I_r) \leftarrow (p_1,I_1)$;
  \FOR{$i=1$ to $M+1$}
   \STATE $(p_t,I_t) \ \leftarrow\ W_i \cdot (p_r,I_r) + B_i$; \COMMENT{Using TM arithmetic}
   \STATE Computing a polynomial approximation $p_{\sigma,i}$ for $\sigma$ w.r.t. the domain $(p_t,I_t)$;
   \STATE Evaluating a conservative remainder $I_{\sigma,i}$ for $p_{\sigma,i}$ w.r.t. the domain $(p_t,I_t)$;
   \STATE $(p_r,I_r) \ \leftarrow\ p_{\sigma,i}(p_t + I_t) + I_{\sigma,i}$;  \COMMENT{Using TM arithmetic}
  \ENDFOR
  \RETURN $(p_r,I_r)$.
 \end{algorithmic}
\end{algorithm}

We give the whole procedure by Algorithm~\ref{algo:nn_output}. In our approach, the polynomial approximation $p_{\sigma,i}$ and its remainder interval $I_{\sigma,i}$ for the vector of activation functions $\sigma$ in the $i$-th layer can be computed in the following two ways.

\noindent\textbf{Taylor approximation.}
When the activation function is differentiable in the range defined by $(p_t,I_t)$. The polynomial $p_{\sigma,i}$ can be computed as the order $k$ Taylor expansion of $\sigma$ (in each of its dimension) at the center of $(p_t,I_t)$, and the remainder is evaluated using interval arithmetic based on the Lagrange remainder form. More details are described elsewhere~\cite{Makino+Berz/2003/Taylor}.

The following theorem states that a TM flowpipe computed by our approach is not only a range overapproximation of a reachable set segment, but also a function overapproximation for the dependency of a reachable state on its initial state. The proof is given in the appendix.

\begin{theorem}\label{thm:state-wise}
 If $\mathcal{F}(\vx_0,\tau)$ is the $i$-th TM flowpipe computed in the $j$-st control step, then for any initial state $\vx_0\in X_0$, the box $\mathcal{F}(\vx_0,\tau)$ contains the actual reachable state $\varphi_\mathcal{N}(\vx_0,(j-1)\delta_c + (i-1)\delta + \tau)$ for all $\tau\in [0,\delta]$.
\end{theorem}

\subsection{Bernstein Approximation}

\noindent\textbf{Bernstein approximation in B{\'e}zier form.}
The use of Bernstein approximation only requires the activation function to be continuous in $(p_t,I_t)$, and can be used not only in more general situations, but also to obtain better polynomial approximations than Taylor expansions (see~\cite{Lorentz/Bernstein}). We first give a general method to obtain a Bernstein overapproximation for an arbitrary continues function, and then present a more accurate approach only for ReLU functions.

\begin{figure}
  \vspace{-0.2cm}
  \begin{center}
    \includegraphics[width=0.8\textwidth]{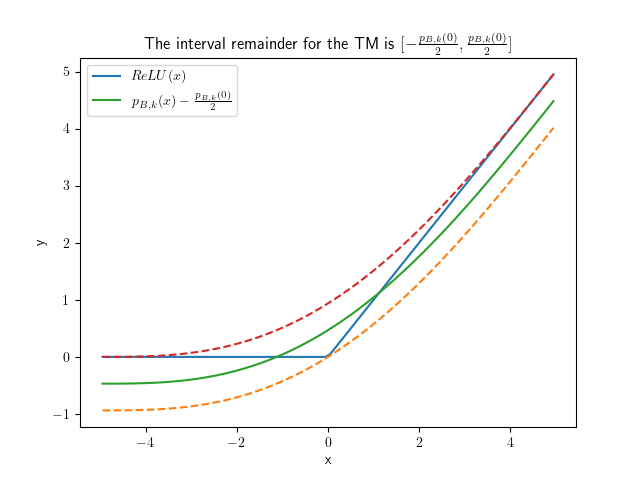}
  \end{center}
  \caption{
  The Taylor model (TM) overapproximation $p(x)+I$ of $\text{ReLU}(x)$ is given by $p(x) = p_{B,k}(x) - \frac{p_{B,k}(0)}{2}$ and $I = [-\frac{p_{B,k}(0)}{2}, \frac{p_{B,k}(0)}{2}])$ where $p_{B,k}(0)$ is the Bernstein polynomial $p_{B,k}(x)$ evaluated at $x=0$. It can be shown that for $x \in [a, b]$ with $a < 0 < b$, the bounds of the interval remainder $I$ are tight for any order-k Bernstein polynomials approximation with $k \geq 1$.
  }
  \label{fig:bp_relu_ub_lb}
  \vspace{-0.5cm}
\end{figure}

\noindent\textit{Bernstein approximation for $\sigma(\vz)$ w.r.t. $\vz \in (p_t,I_t)$.}
Given $(p_t,I_t)$ computed in Line 3, the $j$-th component of the polynomial vector $p_{\sigma,i}$ is the order $k$ Bernstein polynomial of the activation function $\sigma_j$ of the $j$-th neuron. It can be computed as
$p_{\sigma_j,i}(z_j) {=} \sum_{s=0}^{k}\left(\sigma_j(\frac{\bar{Z}_j-\underline{Z}_j}{k}s+\underline{Z}_j)\binom{k}{s}\frac{(Z_j-\underline{Z}_j)^{s}(\bar{Z}_j - z_j)^{k-s}}{(\bar{Z}_j-\underline{Z}_j)^k}\right)$, such that $\bar{Z}_j$ and $\underline{Z}_j$ denote the upper and lower bounds respectively of the range in the $j$-th dimension of $(p_t,I_t)$, and they can be obtained by interval evaluation of the TM.

\noindent\textit{Evaluating the remainder $I_{\sigma,i}$.}
The $j$-th component $I_{\sigma_j,i}$ of $I_{\sigma,i}$ is computed as a conservative remainder for the polynomial $p_{\sigma_j,i}$, and it can be obtain as a symmetric interval $[-\epsilon_j,\epsilon_j]$ such that
\[
 \small
    \epsilon_j {=} \max_{s=1,{\cdots} ,m}\left(\left|p_{\sigma_j,i}(\frac{\overline{Z
    }_j{-}\underline{Z}_j}{m}(s{-}\frac{1}{2}){+}\underline{Z}_j) {-} {\sigma_j}(\frac{\overline{Z}_j-\underline{Z}_j}{m}(s-\frac{1}{2}){+}\underline{Z}_j)\right| {+} L_j{\cdot}\frac{\overline{Z}_j{-}\underline{Z}_j}{m} \right)
\]
wherein $L_j$ is a Lipschitz constant of $\sigma_j$ with the domain $(p_t,I_t)$, and $m$ is the number of samples that are uniformly selected to estimate the remainder. The soundness of the error bound estimation above has been proven in \cite{HuangFLC019} for multivariate Bernstein polynomials. Since univariate Bernstein polynomials, which we use in this paper, is a special case of multivariate Bernstein polynomials, our approach is also sound. A detailed proof is given in the appendix.


\noindent\textbf{Efficient and Tight Error Bound Estimation for ReLU.}
In the general Bernstein overapproximation method, the computed Bernstein polynomial are ``bloated'' by adding a symmetric interval remainder whose radius is the error bound estimated based on samples and a Lipschitz constant. Such a method can be too conservative when the overapproximated function is convex or concave. When a continuous function is convex in an interval domain, its Bernstein approximation is no smaller than it at any point in the domain~\cite{goodman/phillips_1999}. Hence, an order $k$ Bernstein approximation $p(x)$ of a ReLU function $\relu(x)$ with $x\in [a,b]$ is always no smaller than it, and $p(x) + [\varepsilon,0]$ will be a tight overapproxiamtion, such that $\varepsilon = p(0)$. We center the remainder interval at $0$, and a Taylor model for $\relu(x)$ for some $x\in [a,b]$ can be obtained as $p(x) - 0.5\varepsilon + [-0.5\varepsilon,0.5\varepsilon]$. An example is shown in Fig.~\ref{fig:bp_relu_ub_lb}. This approach was first implemented in the POLAR submission to the Artificial Intelligence and Neural Network Control Systems (AINNCS) category in ARCH-COMP 2022~\cite{arch-comp-2022-ainncs}.



\begin{lemma}\label{lem:bernstein_bound}
 Given that $p_k(x)$ is the order $k\geq 1$ Bernstein polynomial of a convex function $f(x)$ with $x\in [a,b]$. For all $x\in [a,b]$, we have that (i) $f(x) \leq p_k(x)$ and (ii) $p_{k+1}(x) \leq p_k(x)$.
\end{lemma}
\begin{proof}
 The Lemma is proved in~\cite{goodman/phillips_1999} for the domain $x \in [0,1]$. However, it also holds on an arbitrary domain $x \in [a,b]$ after we replace the lower and upper bounds in the Bernstein polynomials by $a$ and $b$. \hfill$\Box$
\end{proof}

\begin{corollary}
 If $p(x)$ is the order $k\geq 1$ Bernstein polynomial of $\relu(x)$ with $x\in [a,b]$, then $0\leq \relu(x) \leq p(x)$ for all $x\in [a,b]$.
\end{corollary}

\begin{lemma}
 Given that $p(x)$ is the order $k\geq 1$ Bernstein polynomial of $\relu(x)$ with $x\in [a,b]$ such that $a < 0 < b$, then we have that $p(x) - \relu(x)\leq p(0)$ for all $x\in [a,b]$.
\end{lemma}
\begin{proof}
 Since $\relu(x)$ is convex over the domain, by~\cite{goodman/phillips_1999}, so is $p(x)$. Therefore, the second derivative of $p$ w.r.t. $x$ is non-negative. By evaluating the first derivatives of $p$ at $x = a$ and $x = b$, we have that $\frac{dp}{dx}|_{x = a} \geq 0$ and $\frac{dp}{dx}|_{x = b} \leq 1$. Since the first derivatives of $\relu(x)$ are $0$ and $1$ when $x\in [a,0)$ and $x\in (0,b]$ respectively, $p(a) = \relu(a)$, and $p(b) = \relu(b)$, we have that the function $p(x) - \relu(x)$ monotonically increasing when $x\in [a,0]$ and decreasing when $x\in [0,b]$, hence its maximum value is given by $p(0)$.
 \hfill$\Box$
\end{proof}

\subsection{Selection of Polynomial Approximations}

Since an activation function is univariate, both of its Taylor and Bernstein approximations have a size which is linear in the order $k$. Then we investigate the accuracy produced by both approximation forms. Since the main operation in the TM layer-by-layer propagation framework is the composition of TMs, we study the \emph{preservation of accuracy} for both of the forms under the composition with a given TM. We first define the \emph{Accuracy Preservation Problem}.

When a function $f(\vx)$ is overapproximated by a TM $(p(\vx), I)$ w.r.t. a bounded domain $D$, the approximation quality, i.e., size of the overestimation, is directly reflected by the width of $I$, since $f(\vx) = p(\vx)$ for all $\vx\in D$ when $I$ is zero by the TM definition.
Given two order $k$ TMs $(p_1(\vx), I_1)$ and $(p_2(\vx), I_2)$ which are overapproximations of the same function $f(\vx)$ w.r.t. a bounded domain $D \subset \reals^n$, we use $(p_1(\vx), I_1) \prec_k (p_2(\vx), I_2)$ to denote that the width of $I_1$ is smaller than the width of $I_2$ in all dimensions, i.e., $(p_1(\vx), I_1)$ is a more accurate overapproximation of $f(\vx)$ than $(p_2(\vx), I_2)$.

\noindent\textbf{Accuracy Preservation Problem.}
If both $(p_1(\vx), I_1)$ and $(p_2(\vx), I_2)$ are overapproximations of $f(\vx)$ with $\vx\in D$, and $(p_1(\vx), I_1) \prec_k (p_2(\vx), I_2)$. Given another function $g(\vy)$ which is already overapproximated by a TM $(q(\vy), J)$ whose range is contained in $D$. Then, \emph{does $p_1(q(\vy) + J) + I_1\, \prec_k\, p_2(q(\vy) + J) + I_2$ still hold using order $k$ TM arithmetic?}

We give the following counterexample to show that the answer is \textbf{no}, i.e., although $(p_1(\vx), I_1)$ is more accurate than $(p_2(\vx), I_2)$, the composition $p_1(q(\vy) + J) + I_1$ might not be a better order $k$ overapproximation than $p_2(q(\vy) + J) + I_2$ for the composite function $f\circ g$. Given $p_1 = 0.5 + 0.25 x - 0.02083 x^3$, $I_1 =$ [-7.93e-5, 1.92e-4], and $p_2 = 0.5 + 0.24855 x - 0.004583 x^3$, $I_2 =$ [-2.42e-4, 2.42e-4], which are both TM overapproximations for the sigmoid function $f(x) = \frac{1}{1 + e^{-x}}$ w.r.t. $x\in q(y) + J$ such that $q=0.1 y - 0.1 y^2$, $J = [-0.1,0.1]$, and $y\in [-1,1]$. We have that $(p_1,I_1) \prec_3 (p_2,I_2)$, however after the compositions using order $3$ TM arithmetic, the remainder of $p_1(q(y) + J) + I_1$ is $[-0.0466 , 0.0477]$, while the remainder of $p_2(q(y) + J) + I_2$ is $[-0.0253 , 0.0253]$, and we do not have $(p_1(q(y) + J) + I_1) \prec_3 (p_1(q(y) + J) + I_1)$.

Hence, we integrate an additional step in Algorithm~\ref{algo:nn_output} to replace line 4-6: in each iteration, both of Taylor and Bernstein overapproixmations are computed for each of the activation functions, and we choose the one that produces the smaller remainder interval $I_r$.

\subsection{Symbolic Remainders in Layer-by-Layer Propagation}

We describe the use of symbolic remainders (SR) in the layer-by-layer propagation of computing an NN output TM. The method was originally proposed in~\cite{chen/decomposition} for reducing the overestimation of TM flowpipes in the reachability computation for nonlinear ODEs, we adapt it particularly for reducing the error accumulation in the TM remainders during the layer-by-layer propagation. Unlike the BP technique whose purpose is to obtain tighter TMs for activation functions, the use of SR only aims at reducing the overestimation accumulation in the composition of a sequence of TMs each of which represents the input range of a layer.

\begin{algorithm}[tbp]
 \caption{TM output computation using symbolic remainders, input and output are the same as those in Algorithm~\ref{algo:nn_output}}\label{algo:sym_rem}
 \begin{algorithmic}[1]
  \STATE Setting $\mathcal{Q}$ as an empty array which can keep $M+1$ matrices;
  \STATE Setting $\mathcal{J}$ as an empty array which can keep $M+1$ multidimensional intervals;
  \STATE $\mathbb{J} \leftarrow 0$;
  \FOR{$i=1$ to $M+1$}
   \STATE Computing the composite function $p_{\sigma,i}$ and the remainder interval $I_{\sigma,i}$ using the BP technique;
   \STATE Evaluating $q_i(\vx_0) + J_i$ based on $\mathbb{J}$ and $\mathcal{Q}[1]I_1$; \COMMENT{$\mathcal{Q}[1]I_1 = I_1$ when $i=1$}
   \STATE $\mathbb{J} \leftarrow J_i$;
   \STATE $\Phi_i = Q_i W_i$;
   \FOR{$j=1$ to $i-1$}
    \STATE $ \mathcal{Q}[j] \leftarrow \Phi_i \cdot \mathcal{Q}[j]$;
   \ENDFOR
   \STATE Adding $\Phi_i$ to $\mathcal{Q}$ as the last element;
   \FOR{$j=2$ to $i$}
    \STATE $\mathbb{J} \leftarrow \mathbb{J} + \mathcal{Q}[j] \cdot \mathcal{J}[j-1]$;
   \ENDFOR
   \STATE Adding $J_i$ to $\mathcal{J}$ as the last element;
  \ENDFOR
  \STATE Computing an interval enclosure $I_r$ for $\mathbb{J} + \mathcal{Q}[1]I_1$; \COMMENT{interval evaluation}
  \RETURN $q_{M+1}(\vx_0) + I_r$.
 \end{algorithmic}
\end{algorithm}


Consider the TM composition for computing the output TM of a single layer in  Fig.~\ref{fig:single_layer_composition},
the output TM $p_{\sigma,i}(W_i (p_i(\vx_0) + I_i) + B_i) + I_{\sigma,i}$ equals to $Q_i W_i p_i(\vx_0) + Q_i W_i I_i + Q_i B_i + p_{\sigma,i}^R(W_i (p_i(\vx_0) + I_i) + B_i) + I_{\sigma,i}$ such that $Q_i$ is the matrix of the linear coefficients in $p_{\sigma,i}$, and $p_{\sigma,i}^R$ consists of the terms in $p_{\sigma,i}$ of the degrees $\neq 1$. Therefore, the remainder $I_i$ in the second term can be kept symbolically such that we do not compute $Q_i W_i I_i$ out as an interval but keep its transformation matrix $Q_i W_i$ to the subsequent layers. Given the image $S$ of an interval under a linear mapping, we use $\underline{S}$ to denote that it is kept symbolically, i.e., we keep the interval along with the transformation matrix, and $\overline{S}$ to denote that the image is evaluated as an interval.

Then we present the use of SR in layer-by-layer propagation. Starting from the NN input TM $(p_1(\vx_0),I_1)$, the output TM of the first layer is computed as
\[
 \small
 \underbrace{Q_1 W_1 p_1(\vx_0) + Q_1 B_1 + p_{\sigma,1}^R(W_1 (p_1(\vx_0) + I_1) + B_1) + I_{\sigma,1}}_{q_1(\vx_0) + J_1} + \underline{Q_1 W_1 I_1}
\]
which can be kept in the form of $q_1(\vx_0) + J_1 + \underline{Q_1 W_1 I_1}$. Using it as the input TM of the second layer, we have the following TM
\[
 \small
 \begin{aligned}
   & p_{\sigma,2}(W_2 (q_1(\vx_0) + J_1 + \underline{Q_1 W_1 I_1}) + B_2) + I_{\sigma,2} \\
   = & \underbrace{Q_2 W_2 q_1(\vx_0) + Q_2 B_2 + p_{\sigma,2}^R(W_2 (q_1(\vx_0) + J_1 + \overline{Q_1 W_1 I_1}) + B_2) + I_{\sigma,2}}_{q_2(\vx_0) + J_2} \\
   & + \underline{Q_2 W_2 J_1} + \underline{Q_2 W_2 Q_1 W_1 I_1}
 \end{aligned}
\]
for the output range of the second layer. Therefore the output TM of the $i$-th layer can be obtained as $q_i(\vx_0) + \mathbb{J}_i + \underline{Q_iW_i\cdots Q_1W_1 I_1}$ such that $\mathbb{J}_i = J_i + \underline{Q_iW_iJ_{i-1}} + \underline{Q_iW_iQ_{i-1}W_{i-1}J_{i-2}} + \cdots + \underline{Q_iW_i\cdots Q_2W_2 J_1}$.

We present the SR method by Algorithm~\ref{algo:sym_rem} in which we use two lists: $\mathcal{Q}[j]$ for $Q_iW_i\cdot\cdots\cdot Q_jW_j$ and $\mathcal{J}[j]$ for $\mathbb{J}_j$ to keep the intervals and their linear transformations. The symbolic remainder representation is replaced by its interval enclosure $I_r$ at the end of the algorithm.


\noindent\textbf{Time and space complexity.}
Although Algorithm~\ref{algo:sym_rem} produces TMs with tighter remainders than Algorithm~\ref{algo:nn_output} because of the symbolic interval representations under linear mappings, it requires (1) two extra arrays to keep the intermediate matrices and remainder intervals, (2) two extra inner loops which perform $i-1$ and $i-2$ iterations in the $i$-th outer iteration. The size of $Q_iW_i\cdot\cdots\cdot Q_jW_j$ is determined by the rows in $Q_i$ and the columns in $W_j$, and hence the maximum number of neurons in a layer determines the maximum size of the matrices in $\mathcal{Q}$. Similarly, the maximum dimension of $J_i$ is also bounded by the maximum number of neurons in a layer. Because of the two inner loops, time complexity of Algorithm~\ref{algo:sym_rem} is quadratic in $M$, whereas Algorithm~\ref{algo:nn_output} is linear in $M$.

\section{Experiments}\label{sec:experiments}

In this section, we perform a comprehensive empirical study of POLAR against state-of-the-art (SOTA) techniques. We first demonstrate the performance of POLAR on two examples with high dimensional states and multiple inputs, which are far beyond the ability of current SOTA techniques (Section 4.1). A comprehensive comparison with SOTA over the full benchmarks in \cite{HuangFLC019,ivanov2021veification} is then given (Section 4.2). Finally, we present additional ablation studies, scalability analysis, and the ability to handle discrete-time systems (Section 4.3).

All our experiments were run on a machine with 6-core 2.20 GHz Intel Core i7 and 16GB of RAM. POLAR is implemented with C++. We present the results for POLAR, Verisig 2.0 and Sherlock using a single core without parallelization.
The results of ReachNN* were computed on the same machine with the aid of GPU acceleration on an Nvidia GeForce RTX 1050Ti GPU.

\noindent\textbf{State-of-the-art tools.} We compare with SOTA tools in the NNCS reachability analysis literature, including Sherlock \cite{DuttaCS19} (only works for ReLU), Verisig 2.0 \cite{ivanov2021veification} (only works for sigmoid and tanh),
NNV \cite{TranYLMNXBJ20},
and ReachNN*\cite{FanHCL020}\footnote{The results of ReachNN* are based on GPU acceleration.}.




\begin{figure}[h!]
    \centering
    \subfloat{
	\includegraphics[width=.32\textwidth]{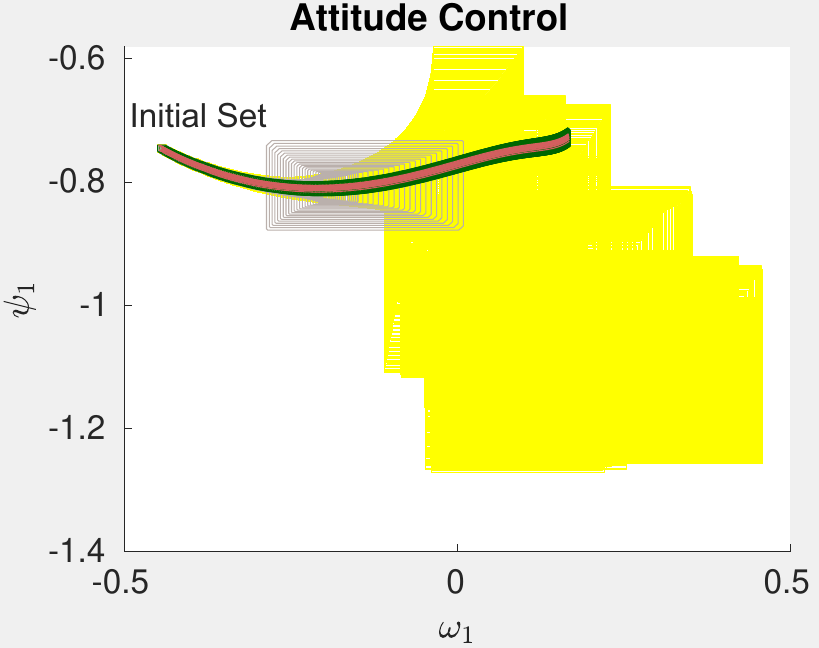}}
	\hfill
	\subfloat{
	\includegraphics[width=.32\textwidth]{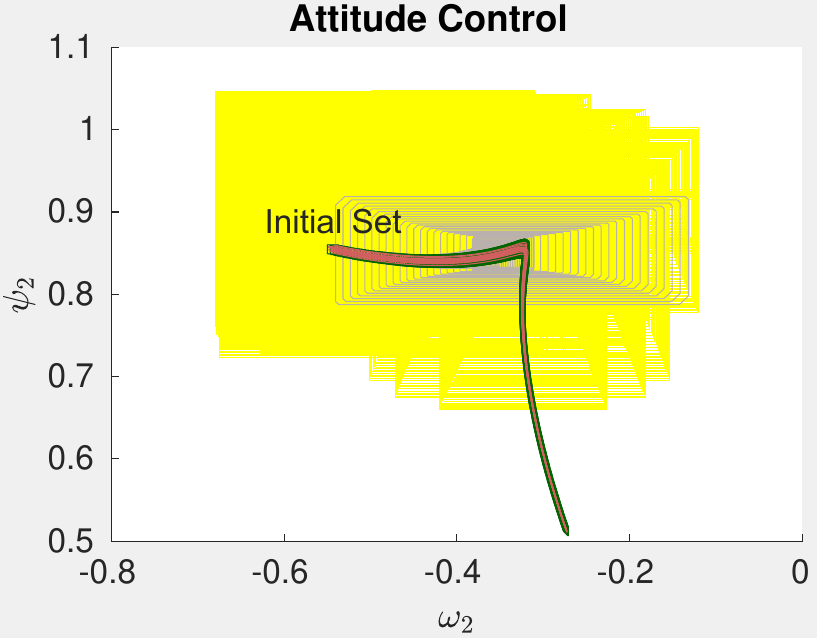}}
	\hfill
	\subfloat{
	\includegraphics[width=.32\textwidth]{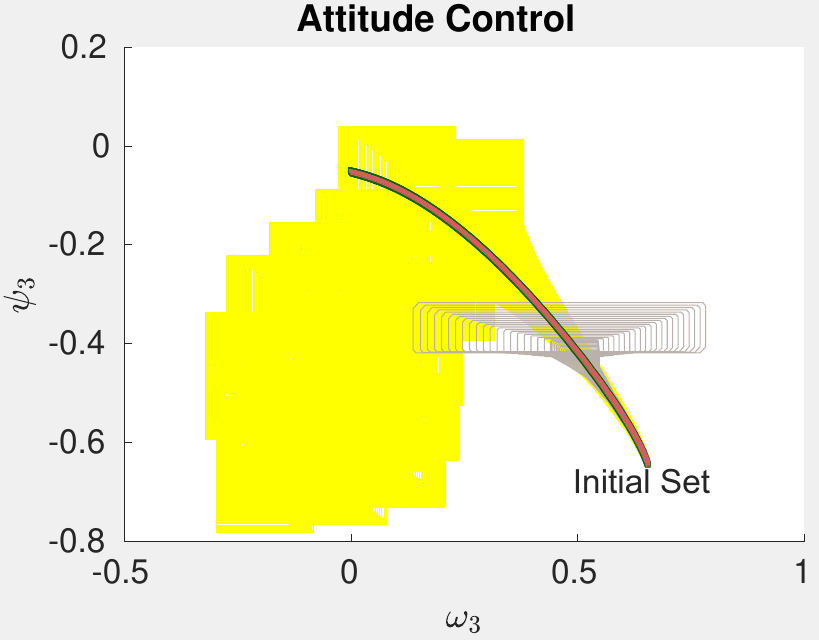}}
    \caption{Comparison between reachable sets of the 6-dimensional attitude control benchmark produced by POLAR (dark green), Verisig 2.0 (gray) and NNV (yellow). The red curves are simulated trajectories.} 
    \label{fig:attitude-control}
\end{figure}

\begin{figure}[h!]
    \centering
    \subfloat[QUAD\label{subfig:quad}]{
    \includegraphics[width=0.45\columnwidth]{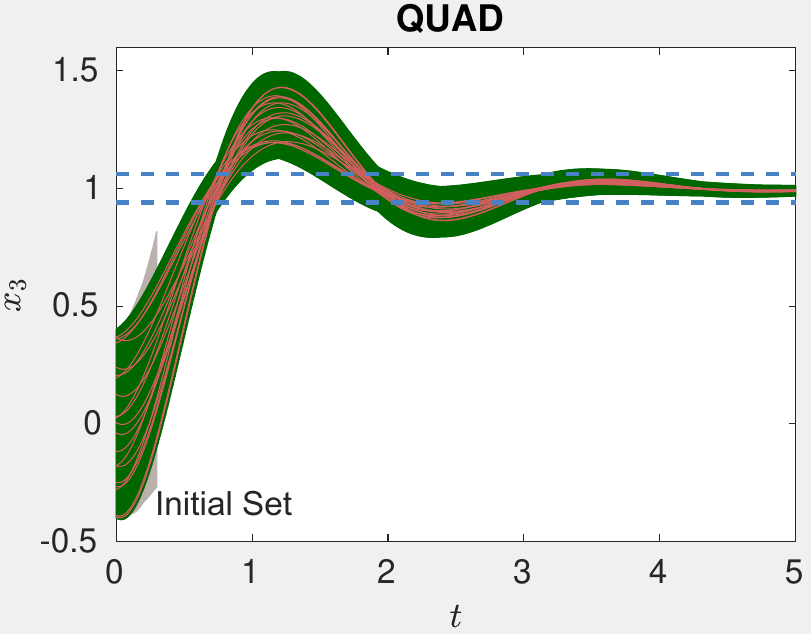}}
    \hfill
    \subfloat[Mountain Car \label{subfig:mc}]{\includegraphics[width=0.45\columnwidth]{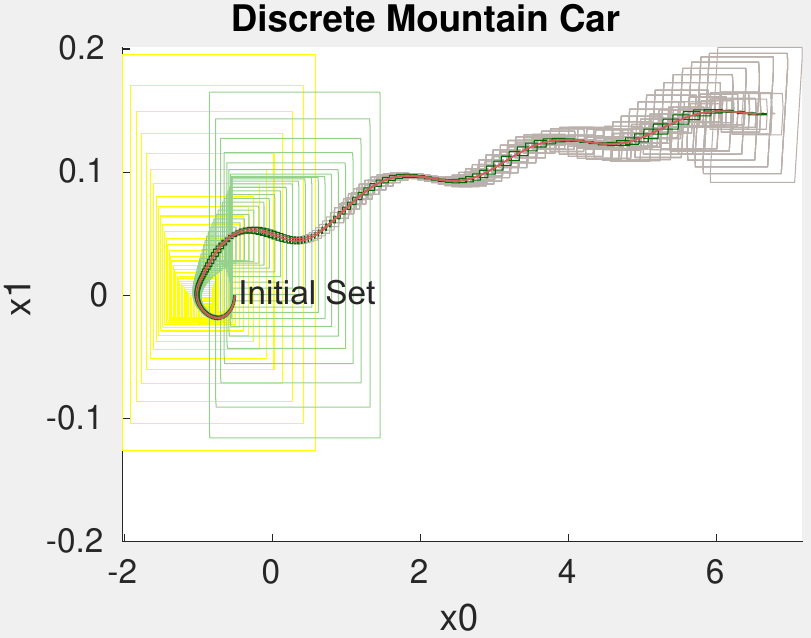}}
    \caption{(a) Results of QUAD. POLAR for 50 steps (dark green sets), Verisig 2.0 for 3 steps (grey sets), and simulation traces for 50 steps (red curves). It took POLAR 1271 seconds to compute the flowpipes for 50 steps. On the other hand, it took Verisig 2.0 more than 5 hours to compute the flowpipes for the first 3 steps, and 
    at the $4^{\text{th}}$ step, the remainders of the TM computed by Verisig 2.0 for the outputs of the neural-network controller already exploded to $10^{15}$. NNV crashed with out-of-memory errors when computing the $1^{\text{st}}$ step.
    (b) Results of Mountain Car. POLAR for 150 steps (dark green sets), Verisig 2.0 for 150 steps (grey sets), ReachNN* for 90 steps (light green sets), NNV for 65 steps, and simulation traces for 150 steps (red curves). }
    \label{fig:quad}
\end{figure}

\subsection{High Dimensional Case Studies: Attitude Control \& QUAD.} 

We consider an attitude control of a rigid body with 6 states and 3 control inputs~\cite{prajna2004nonlinear}, and quadrotor (QUAD) with 12 states and 3 control inputs~\cite{beard2008quadrotor} to evaluate the performance of POLAR on difficult problems. The complexity of these two example lies in the combination of the numbers of the state variables and control inputs. For each example, we trained a sigmoid neural-network controller and compare POLAR with Verisig 2.0 and NNV. The detailed setting of these two examples can be found in the Appendix.

The result for the attitude control benchmark is shown in Figure \ref{fig:attitude-control}, and the result for the QUAD benchmark is shown in Figure \ref{subfig:quad}. In the attitude control benchmark, POLAR computed the TM flowpipes for 30 control steps in 201 seconds. From Figure \ref{fig:attitude-control},
We can observe that the flowpipes computed by POLAR are tight w.r.t. the simulated traces.
As a comparison, although Verisig 2.0 \cite{ivanov2021veification} can handle this system in theory, its remainder exploded very quickly and the tool crashed after only a few steps. NNV computed flowpipes for 25 steps by doing extensive splittings on the state space and crashed with out-of-memory errors. In the QUAD benchmark, POLAR computed the TM flowpipes for 50 control steps in 1271 seconds, while Verisig 2.0 and NNV took hours to compute flowpipes just for the first few steps.

\begin{table}[t] 
\center
\caption{$V$: number of state variables, $\sigma$: activation functions, $M$: number of hidden layers, $n$: number of neurons in each hidden layer. For each approach (POLAR, ReachNN*, Sherlock, Verisig 2.0), we give the runtime in seconds if it successfully verifies the property.
`Unknown': the property could not be verified. `--': the approach cannot be applied due to the type of $\sigma$.
}
\begin{adjustbox}{max width=\columnwidth, max height=1.8in}
\begin{threeparttable}
\begin{tabular}{|c|c|c|c|c|c|c|c|c|}
\hline
\multirow{2}{*}{\#} &  \multirow{2}{*}{V}  & \multicolumn{3}{c|}{NN Controller} &
\multirow{2}{*}{\textbf{POLAR}}  & \multirow{1}{*}{ReachNN*}  & \multirow{1}{*}{Sherlock}  & \multirow{1}{*}{Verisig 2.0}   \\ \cline{3-5}    
                &        &        $\sigma$        & M    & n &   
                & \cite{FanHCL020} & \cite{DuttaCS19} & \cite{ivanov2021veification}  \\ \hline
\multirow{4}{*}{1} & \multirow{4}{*}{2} & ReLU       &   2   &  20 &  \textbf{13}   & 29       &  42 & --   \\ \cline{3-9}     
                   &    &      sigmoid    &   2   &  20 &      \textbf{23}     &   73     &  -- &   70  \\ \cline{3-9}     
                   &    &      tanh       &   2   &  20 & \textbf{25}          & Unknown         &  -- &    70  \\ \cline{3-9}     
                   &    &      ReLU+tanh       &   2   &  20 &   \textbf{13}         &   72     &  -- & --    \\ \hline
\multirow{4}{*}{2} & \multirow{4}{*}{2} & ReLU       &   2   &  20 &      \textbf{1}    &    5   &  3 & --   \\ \cline{3-9}     
                   &    &      sigmoid    &   2   &  20 & \textbf{10}        &   12   &  -- &  11  \\ \cline{3-9}     
                   &    &      tanh       &   2   &  20 &  \textbf{3}       &   75    &  --   &  Unknown   \\
                   \cline{3-9}     
                   &    &      ReLU+tanh           &   2   &  20  &      \textbf{2}    &   Unknown    &  -- & --   \\ \hline 
\multirow{4}{*}{3} & \multirow{4}{*}{2} & ReLU       &   2   &  20 &      \textbf{13}    &   89      &  143 & --    \\ \cline{3-9}     
                   &    &      sigmoid    &   2   &  20 &   \textbf{37}    &   141    &  -- &   64   \\ \cline{3-9}     
                   &    &      tanh       &   2   &  20 &  \textbf{38}      &   139   &  -- & 54   \\ \cline{3-9}     
                   &    &      ReLU+sigmoid       &   2   &  20  &  \textbf{14}      & 146      &  -- & --  \\   \hline
\multirow{4}{*}{4} & \multirow{4}{*}{3} & ReLU       &  2    &  20 &      \textbf{1}    & 9       &  21 & --  \\                                 \cline{3-9}     
                   &    &      sigmoid    &  2    &   20 &    \textbf{4}   &   22     &  -- &  15  \\ \cline{3-9}     
                   &    &      tanh    &  2    &   20  & \textbf{4}      &  22     &  -- & 14 \\ \cline{3-9}     
                   &    &      ReLU+tanh       &   2   &  20   &    \textbf{1}    & 12      &  -- & -- \\   \hline
\multirow{4}{*}{5} & \multirow{4}{*}{3}  & ReLU       &   3   &  100 &      \textbf{5}     &   117     &  15 & --    \\ \cline{3-9}     
                   &    &      sigmoid    &  3    &   100     &  \textbf{25}   &    41   &  -- &  280 \\ \cline{3-9}     
                   &    &      tanh       &   3   &  100 &  \textbf{31}       &    Unknown   &  -- &    265  \\ \cline{3-9}     
                   &    &      ReLU+tanh       &   3   &  100  &  \textbf{5}      & Unknown   & -- & --  \\
                   \hline
\multirow{4}{*}{6} & \multirow{4}{*}{4}  & ReLU       &    3  &  20 &      \textbf{19}    &     1130\tnote{1}   &  35 & --   \\ \cline{3-9}     
                &    &      sigmoid    &  3    &   20  &   \textbf{30}      &    13350\tnote{1}       &  -- &  121 \\ \cline{3-9}     
                   &    &      tanh       &   3   &  20 &   \textbf{32}       & 2416\tnote{1}  &  -- &    100   \\ \cline{3-9}     
                   &    &      ReLU+tanh       &   3   &  20  &   \textbf{20}     &     1413\tnote{1}   &  -- & --  \\   \hline
ACC & 6 & tanh & 3 & 20 & \textbf{312} & Unknown & -- & 5045  \\ \hline
QMPC & 6 & tanh & 2 & 20 & \textbf{61} & --\tnote{2} & -- & 1065  \\ \hline
Attitude Control & 6 & sigmoid & 3 & 64 & \textbf{194} & --\tnote{2} & -- & Unknown \\ \hline
QUAD & 12 & sigmoid & 3 & 64 & \textbf{1271} & --\tnote{2} & -- & Unknown \\
\hline
\end{tabular}
\begin{tablenotes}
    \item[1] ReachNN* runs out of memory for this example, we then use another machine with 128G memory to obtain the runtime result. 
    \item[2] This example has multi-dimensional control inputs. ReachNN* only supports NN controllers that produce single-dimensional control inputs.
\end{tablenotes}
\end{threeparttable}
\end{adjustbox}
\label{tab:results}
\end{table}

\subsection{Comparison over A Full Set of Benchmarks}

We compare POLAR with the SOTA tools mentioned previously, including Sherlock, Verisig 2.0, NNV, and ReachNN* over the full benchmarks in \cite{HuangFLC019,ivanov2021veification}.
We refer to \cite{HuangFLC019,ivanov2021veification} for more details of these benchmarks.
The results are presented in Table \ref{tab:results} where NNV is not included since we were not able to successfully use it to prove any of the benchmarks likely because it is designed for linear systems. 
Similar results for NNV are also observed in \cite{ivanov2021veification}.
We can see that POLAR successfully verifies all the cases
and the runtime
is
\textbf{on average 8x and up to 71x faster}\footnote{These are lower bounds on the improvements since other tools terminated early for certain settings due to explosion of their computed flowpipes.} compared with the tool with the second best efficiency. The "Unknown" verification results either indicate the  overapproximation of reachable set were too large for verifying the safety property or the tool terminated early due to an explosion of the overapproximation. POLAR achieves the best performance among all the tools (visualizations and detailed comparisons of the reachable sets can be found in the Appendix).

\subsection{Discussion}

POLAR demonstrates substantial performance improvement over existing tools. In this section, we seek to further explore the capability of POLAR. We conduct several experiments for the QUAD benchmark to better understand the limitation and scalability of POLAR. We also include a mountain car example to show that POLAR is able to handle discrete-time systems.

\noindent\textbf{Ablation Studies.} To explore the impact of the two proposed techniques, namely Bernstein polynomial (BP) and symbolic remainder (SR) on the overall performance, we conduct a series of experiments on the QUAD benchmark with different configurations. Table \ref{tab:quad_compair} shows the performance of POLAR with and without the proposed techniques SR and BP in the NN propagation: 1) TM: only TM arithmetic is used; 2) TM+SR: SR is used with TM arithmetic; 3) BP is used with TM arithmetic; and 4) Both BP and SR are used with TM arithmetic. Based on the results, we can observe that SR significantly improves the accuracy of the reachable set overapproximation.
Finally, the combination of basic TM with BP and SP not only achieves the best accuracy, but also is the most efficient.
While the additional BP and SR operations can incur runtime overhead compared with basic TM, they help to produce a tighter overestimation and thus reduce the state space being explored during reachability analysis. As a result, the overall performance including runtime is better.

The following further observations can be obtained from Table~\ref{tab:quad_compair}. (i) Both of the independent use of BP and SR techniques significantly improves the performance of reachable set overapproximations. (ii) When the BP technique is used, Bernstein approximation is often not used on activation functions, but the few times for which they are used  significantly improve the accuracy. The reason of having this phenomenon is that Taylor and Bernstein approximations are similarly accurate in approximating activation functions with small domain. However, the Lagrange form-based remainder evaluation in Taylor polynomials performs better than the sample-based remainder evaluation in Bernstein polynomials in those cases. It can also be seen that for each $X_0$, the use of Bernstein approximation becomes more frequent when the TMs has larger remainders. (iii) When both BP and SR techniques are used, the approach produces the tightest TMs compared with the other columns in the table even though the use Bernstein approximation is less often. The reason is that the remainders of the TMs are already well-limited and most of the activation functions handled in the reachability computation are with a ``small'' TM domain.

\begin{table} [!t] 
\center
\caption{Ablation Studies for POLAR on the QUAD benchmark. We compare the width of TM remainder on $x_3$ at the 50th step under different settings. For settings with BP, we also list the percentage of times where BP is used among 9600 neurons. If a setting cannot compute flowpipes for all 50 steps, it is marked as \emph{Unknown}. $X_0$ is the radius of the initial set. $k$ is the order of the TM.
}\label{tab:experiment}
\setlength\tabcolsep{4.0pt}
\begin{adjustbox}{width=\columnwidth}
\begin{tabular}{|c|c|cc|cc|ccc|ccc|}
\hline
\multirow{2}{*}{$X_0$} & \multirow{2}{*}{$k$} & \multicolumn{2}{c|}{TM} & \multicolumn{2}{c|}{TM+SR} & \multicolumn{3}{c|}{TM+BP} & \multicolumn{3}{c|}{TM+BP+SR} \\ \cline{3-12} 
 &  & \multicolumn{1}{c|}{Width} & Time (s) & \multicolumn{1}{c|}{Width} & Time (s) & \multicolumn{1}{c|}{Width} & \multicolumn{1}{c|}{Time (s)} & BP \% & \multicolumn{1}{c|}{Width} & \multicolumn{1}{c|}{Time (s)} & BP \% \\ \hline
\multirow{3}{*}{0.05} & 2 & \multicolumn{1}{c|}{7.5e-04} & 229 & \multicolumn{1}{c|}{1.3e-04} & 233 & \multicolumn{1}{c|}{6.8e-04} & \multicolumn{1}{c|}{228} & 5.79\% & \multicolumn{1}{c|}{1.2e-04} & \multicolumn{1}{c|}{231} & 1.34\% \\ \cline{2-12} 
 & 3 & \multicolumn{1}{c|}{5.2e-04} & 273 & \multicolumn{1}{c|}{6.5e-05} & 251 & \multicolumn{1}{c|}{5.0e-04} & \multicolumn{1}{c|}{274} & 3.62\% & \multicolumn{1}{c|}{6.5e-05} & \multicolumn{1}{c|}{251} & 0\% \\ \cline{2-12} 
 & 4 & \multicolumn{1}{c|}{4.9.e-04} & 332 & \multicolumn{1}{c|}{6.2e-05} & 270 & \multicolumn{1}{c|}{4.7e-04} & \multicolumn{1}{c|}{336} & 3.57\% & \multicolumn{1}{c|}{6.2e-05} & \multicolumn{1}{c|}{270} & 0\% \\ \hline
\multirow{3}{*}{0.1} & 2 & \multicolumn{1}{c|}{\emph{Unknown}} & -- & \multicolumn{1}{c|}{2.3e-03} & 319 & \multicolumn{1}{c|}{1.0e-02} & \multicolumn{1}{c|}{325} & 9.68\% & \multicolumn{1}{c|}{1.1e-03} & \multicolumn{1}{c|}{289} & 4.80\% \\ \cline{2-12} 
 & 3 & \multicolumn{1}{c|}{1.8e-03} & 352 & \multicolumn{1}{c|}{2.2e-04} & 287 & \multicolumn{1}{c|}{1.7e-03} & \multicolumn{1}{c|}{349} & 6.85\% & \multicolumn{1}{c|}{2.2e-04} & \multicolumn{1}{c|}{287} & 0\% \\ \cline{2-12} 
 & 4 & \multicolumn{1}{c|}{1.6e-03} & 431 & \multicolumn{1}{c|}{1.9e-04} & 304 & \multicolumn{1}{c|}{1.5e-03} & \multicolumn{1}{c|}{427} & 6.70\% & \multicolumn{1}{c|}{1.9e-04} & \multicolumn{1}{c|}{304} & 0\% \\ \hline
\multirow{3}{*}{0.2} & 2 & \multicolumn{1}{c|}{\emph{Unknown}} & -- & \multicolumn{1}{c|}{\emph{Unknown}} & -- & \multicolumn{1}{c|}{\emph{Unknown}} & \multicolumn{1}{c|}{--} & -- & \multicolumn{1}{c|}{\emph{Unknown}} & \multicolumn{1}{c|}{--} & -- \\ \cline{2-12} 
 & 3 & \multicolumn{1}{c|}{9.0e-03} & 721 & \multicolumn{1}{c|}{1.9e-03} & 412 & \multicolumn{1}{c|}{7.8e-03} & \multicolumn{1}{c|}{670} & 4.03\% & \multicolumn{1}{c|}{1.6e-03} & \multicolumn{1}{c|}{394} & 0.77\% \\ \cline{2-12} 
 & 4 & \multicolumn{1}{c|}{5.0e-03} & 761 & \multicolumn{1}{c|}{9.2e-04} & 403 & \multicolumn{1}{c|}{4.7e-03} & \multicolumn{1}{c|}{728} & 4.38\% & \multicolumn{1}{c|}{8.1e-04} & \multicolumn{1}{c|}{396} & 0.07\% \\ \hline
\multirow{3}{*}{0.4} & 2 & \multicolumn{1}{c|}{\emph{Unknown}} & -- & \multicolumn{1}{c|}{\emph{Unknown}} & -- & \multicolumn{1}{c|}{\emph{Unknown}} & \multicolumn{1}{c|}{--} & -- & \multicolumn{1}{c|}{\emph{Unknown}} & \multicolumn{1}{c|}{--} & -- \\ \cline{2-12} 
 & 3 & \multicolumn{1}{c|}{\emph{Unknown}} & -- & \multicolumn{1}{c|}{\emph{Unknown}} & -- & \multicolumn{1}{c|}{\emph{Unknown}} & \multicolumn{1}{c|}{--} & -- & \multicolumn{1}{c|}{\emph{Unknown}} & \multicolumn{1}{c|}{--} & -- \\ \cline{2-12} 
 & 4 & \multicolumn{1}{c|}{\emph{Unknown}} & -- & \multicolumn{1}{c|}{\emph{Unknown}} & -- & \multicolumn{1}{c|}{\emph{Unknown}} & \multicolumn{1}{c|}{--} & -- & \multicolumn{1}{c|}{3.7e-02} & \multicolumn{1}{c|}{1271} & 3.25\% \\ \hline
\end{tabular}
\end{adjustbox}
\label{tab:quad_compair}
\end{table}

\begin{figure}[tb!]
    \centering
	\includegraphics[height = 4cm]{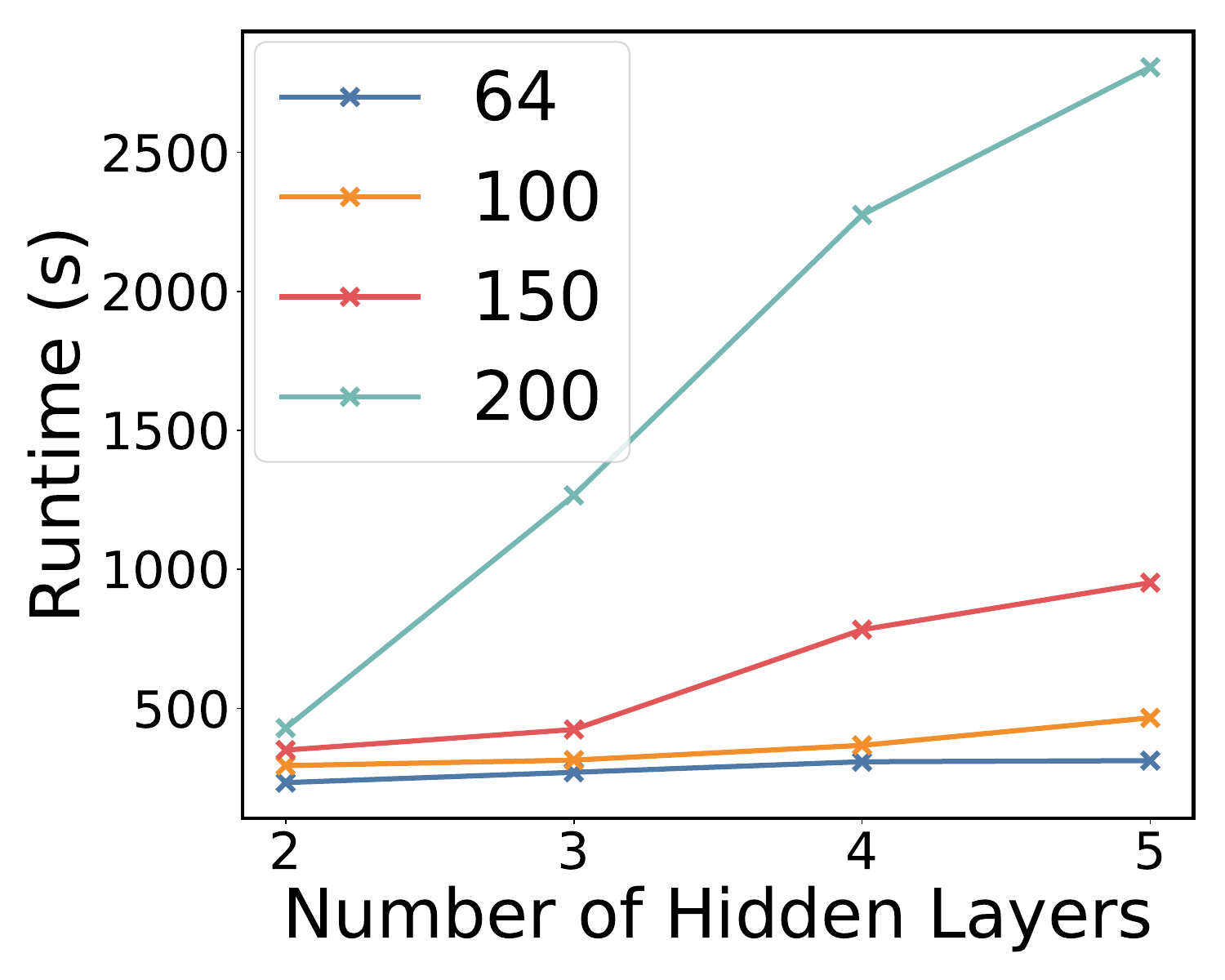}
	\hfill
	\includegraphics[height = 4cm]{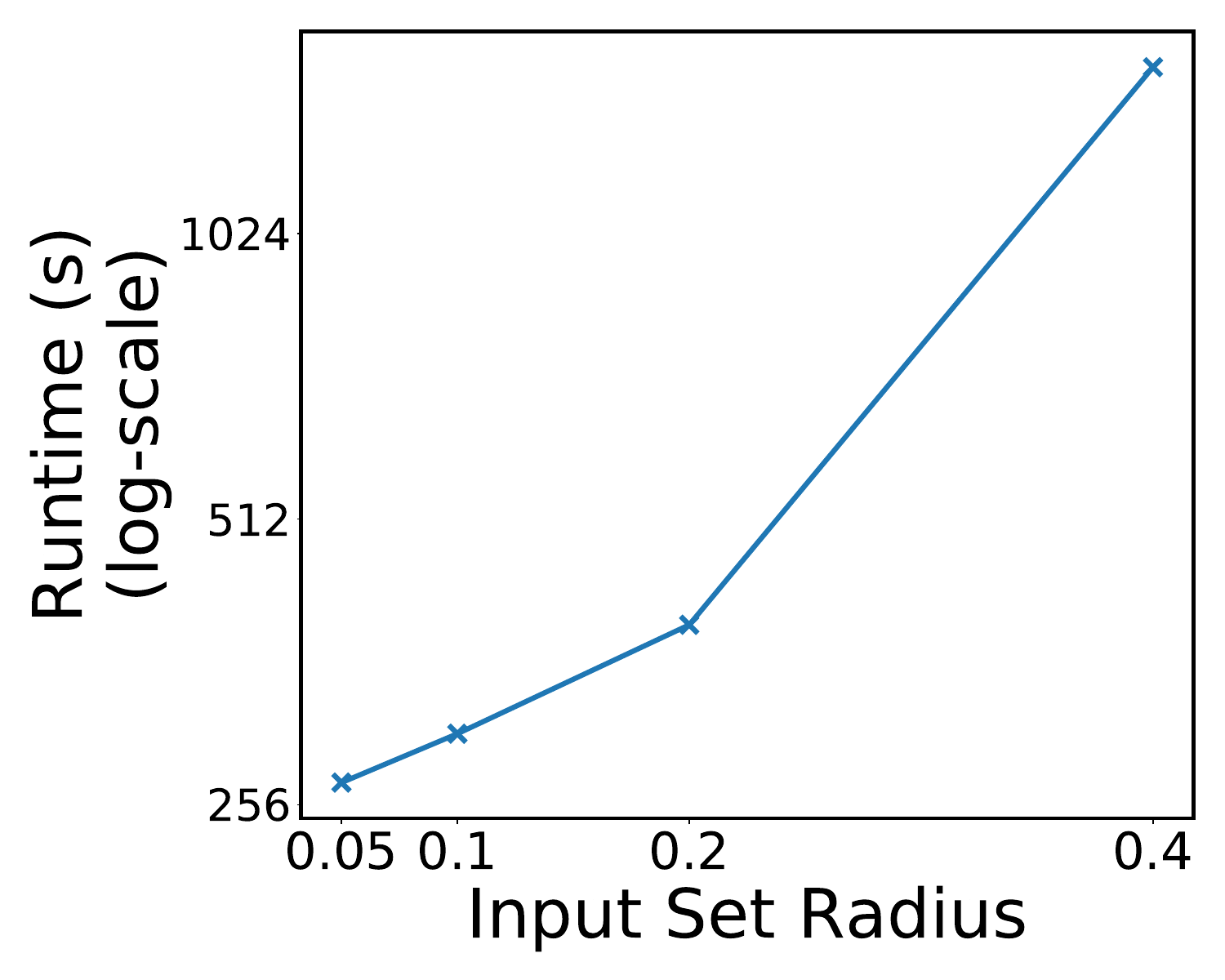}
    \caption{Scalability analysis for POLAR on the QUAD benchmark. We present the runtime of QUAD for 50 steps reachability analysis. Under all settings, POLAR can verify that the system reaches the target set at the 50th step. Left figure: Runtime on different neural network architectures with the input set radius as 0.05. We study neural-network controllers with different number of layers (2, 3, 4, 5) and neurons (64, 100, 150, 200). Right figure: Runtime on the different input set radius of the QUAD benchmark. We use the same network in Figure~\ref{fig:quad} which has 3 hidden layers with 64 neurons in each layer.} 
    \label{fig:scalability}
\vspace{-5mm}
\end{figure}

\noindent\textbf{Scalability Analysis.}  
Table \ref{tab:results} shows that POLAR can handle much larger NNCSs compared with the current SOTA. 
To better understand the scalability of POLAR,
we further conduct scalability analysis 
on the size of the NN controller and the width of the initial set using the QUAD benchmark.
The experiment results in Figure~\ref{fig:scalability} for the neural networks with different widths and depths show that POLAR scales well on the number of layers and the number of neurons in each layer in the NN controller. On the other hand, the time cost grows rapidly when the width of the initial set becomes larger. Such a phenomenon already exists in the literature for reachability analysis of ODE systems~\cite{Chen/2015/phd}. The reason for this is that when the initial set is larger, it is more difficult to track the state dependencies and requires keeping more terms in a TM flowpipe.

\noindent\textbf{Discrete-time NNCS.} Finally, we use Mountain car, a common benchmark in Reinforcement Learning literature, to show that POLAR also works on discrete-time systems. The detailed setting can be found in the Appendix \ref{app_sec:mc}. The comparison with Verisig 2.0, ReachNN* and NNV is shown in Figure \ref{subfig:mc}. POLAR also outperforms these tools substantially for this example.

\section{Conclusion}

In this paper, we propose POLAR, a polynomial arithmetic framework, which integrates TM flowpipe construction, Bernstein overapproximation, and symbolic remainder method to efficiently compute reachable set overapproximations for NNCS. Empirical comparison over a suite of benchmarks shows that POLAR performs significantly better than SOTAs in terms of both computation efficiency and tightness of reachable set estimation. 



\bibliography{bibs/chao,bibs/chen,bibs/jiameng}

\begin{thebibliography}{10}
\providecommand{\url}[1]{\texttt{#1}}
\providecommand{\urlprefix}{URL }
\providecommand{\doi}[1]{https://doi.org/#1}

\bibitem{Althoff/2015/CORA}
Althoff, M.: An introduction to {CORA} 2015. In: International Workshop on
  Applied veRification for Continuous and Hybrid Systems (ARCH). EPiC Series in
  Computing, vol.~34, pp. 120--151 (2015)

\bibitem{Alur+Dill/1994/timed_automata}
Alur, R., Dill, D.L.: A theory of timed automata. Theoretical computer science
  \textbf{126}(2),  183--235 (1994)

\bibitem{beard2008quadrotor}
Beard, R.: Quadrotor dynamics and control rev 0.1. Faculty Publications  (2008)

\bibitem{Berz+Makino/1998/Verified}
Berz, M., Makino, K.: Verified integration of {ODE}s and flows using
  differential algebraic methods on high-order {T}aylor models. Reliable
  computing  \textbf{4},  361--369 (1998)

\bibitem{Chen/2015/phd}
Chen, X.: Reachability Analysis of Non-Linear Hybrid Systems Using Taylor
  Models. Ph.D. thesis, RWTH Aachen University (2015)

\bibitem{Chen+/2013/flowstar}
Chen, X., {\'{A}}brah{\'{a}}m, E., Sankaranarayanan, S.: Flow*: An analyzer for
  non-linear hybrid systems. In: Proc. of CAV'13. LNCS, vol.~8044, pp. 258--263
  (2013)

\bibitem{chen/decomposition}
Chen, X., Sankaranarayanan, S.: Decomposed reachability analysis for nonlinear
  systems. In: Proc. of RTSS'16. pp. 13--24 (2016)

\bibitem{DuttaCS19}
Dutta, S., Chen, X., Sankaranarayanan, S.: Reachability analysis for neural
  feedback systems using regressive polynomial rule inference. In: Proc. of
  HSCC'19. pp. 157--168. {ACM} (2019)

\bibitem{FanHCL020}
Fan, J., Huang, C., Chen, X., Li, W., Zhu, Q.: {ReachNN*}: {A} tool for
  reachability analysis of neural-network controlled systems. In: Proceedings
  of International Symposium on Automated Technology for Verification and
  Analysis (ATVA). LNCS, vol. 12302, pp. 537--542. Springer (2020)

\bibitem{Frehse+/2011/SpaceEx}
Frehse, G., Guernic, C.L., Donz{\'{e}}, A., Cotton, S., Ray, R., Lebeltel, O.,
  Ripado, R., Girard, A., Dang, T., Maler, O.: Spaceex: Scalable verification
  of hybrid systems. In: Proceedings of International Conference on Computer
  Aided Verification (CAV). Lecture Notes in Computer Science, vol.~6806, pp.
  379--395 (2011)

\bibitem{goodman/phillips_1999}
Goodman, T.N.T., Oruç, H., Phillips, G.M.: Convexity and generalized bernstein
  polynomials. Proceedings of the Edinburgh Mathematical Society
  \textbf{42}(1),  179–190 (1999)

\bibitem{HuangFLC019}
Huang, C., Fan, J., Li, W., Chen, X., Zhu, Q.: {ReachNN}: Reachability analysis
  of neural-network controlled systems. {ACM} Trans. Embed. Comput. Syst.
  \textbf{18}(5s),  106:1--106:22 (2019)

\bibitem{HuangKWW17}
Huang, X., Kwiatkowska, M., Wang, S., Wu, M.: Safety verification of deep
  neural networks. In: Proc. of CAV'17. LNCS, vol. 10426, pp. 3--29. Springer
  (2017)

\bibitem{ivanov2021veification}
Ivanov, R., Carpenter, T., Weimer, J., Alur, R., Pappas, G.J., Lee, I.: Verisig
  2.0: Verification of neural network controllers using taylor model
  preconditioning. In: Proc. of CAV'21. LNCS, vol. 12759, pp. 249--262.
  Springer (2021)

\bibitem{IvanovCWAPL21}
Ivanov, R., Carpenter, T.J., Weimer, J., Alur, R., Pappas, G.J., Lee, I.:
  Verifying the safety of autonomous systems with neural network controllers.
  {ACM} Trans. Embed. Comput. Syst.  \textbf{20}(1),  7:1--7:26 (2021)

\bibitem{IvanovWAPL19}
Ivanov, R., Weimer, J., Alur, R., Pappas, G.J., Lee, I.: Verisig: verifying
  safety properties of hybrid systems with neural network controllers. In:
  Proc. of HSCC'18. pp. 169--178. {ACM} (2019)

\bibitem{Jaulin+/2001/applied_interval_analysis}
Jaulin, L., Kieffer, M., Didrit, O., Walter, {\'E}.: Interval analysis. In:
  Applied Interval Analysis. Springer (2001)

\bibitem{KatzBDJK17}
Katz, G., Barrett, C.W., Dill, D.L., Julian, K., Kochenderfer, M.J.: Reluplex:
  An efficient {SMT} solver for verifying deep neural networks. In: Proc. of
  CAV'17. LNCS, vol. 10426, pp. 97--117. Springer (2017)

\bibitem{levine2016end}
Levine, S., Finn, C., Darrell, T., Abbeel, P.: End-to-end training of deep
  visuomotor policies. The Journal of Machine Learning Research
  \textbf{17}(1),  1334--1373 (2016)

\bibitem{arch-comp-2022-ainncs}
Lopez, D.M., Althoff, M., Benet, L., Chen, X., Fan, J., Forets, M., Huang, C.,
  Johnson, T.T., Ladner, T., Li, W., Schilling, C., Zhu, Q.: Arch-comp22
  category report: Artificial intelligence and neural network control systems
  (ainncs) for continuous and hybrid systems plants. In: Frehse, G., Althoff,
  M., Schoitsch, E., Guiochet, J. (eds.) Proceedings of 9th International
  Workshop on Applied Verification of Continuous and Hybrid Systems (ARCH22).
  EPiC Series in Computing, vol.~90, pp. 142--184. EasyChair (2022).
  \doi{10.29007/wfgr}, \url{https://easychair.org/publications/paper/C1J8}

\bibitem{Lorentz/Bernstein}
Lorentz, G.G.: Bernstein Polynomials. American Mathematical Society (2013)

\bibitem{lygeros1999controllers}
Lygeros, J., Tomlin, C.J., Sastry, S.: Controllers for reachability
  specifications for hybrid systems. Automatica  \textbf{35}(3),  349--370
  (1999)

\bibitem{Makino+Berz/2003/Taylor}
Makino, K., Berz, M.: {T}aylor models and other validated functional inclusion
  methods. International Journal of Pure and Applied Mathematics
  \textbf{4}(4),  379--456 (2003)

\bibitem{Meiss/2007/Differential}
Meiss, J.D.: Differential Dynamical Systems. SIAM publishers (2007)

\bibitem{mnih2015human}
Mnih, V., Kavukcuoglu, K., Silver, D., Rusu, A.A., Veness, J., Bellemare, M.G.,
  Graves, A., Riedmiller, M., Fidjeland, A.K., Ostrovski, G., et~al.:
  Human-level control through deep reinforcement learning. Nature
  \textbf{518}(7540),  529--533 (2015)

\bibitem{Moore+Others/2009/Interval}
Moore, R.E., Kearfott, R.B., Cloud, M.J.: Introduction to Interval Analysis.
  SIAM (2009)

\bibitem{pan2018agile}
Pan, Y., Cheng, C., Saigol, K., Lee, K., Yan, X., Theodorou, E.A., Boots, B.:
  Agile autonomous driving using end-to-end deep imitation learning. In: Proc.
  of RSS'18 (2018)

\bibitem{prajna2004safety}
Prajna, S., Jadbabaie, A.: Safety verification of hybrid systems using barrier
  certificates. In: HSCC. pp. 477--492. Springer (2004)

\bibitem{prajna2004nonlinear}
Prajna, S., Parrilo, P.A., Rantzer, A.: Nonlinear control synthesis by convex
  optimization. IEEE Transactions on Automatic Control  \textbf{49}(2),
  310--314 (2004)

\bibitem{SinghGPV19}
Singh, G., Ganvir, R., P{\"{u}}schel, M., Vechev, M.T.: Beyond the single
  neuron convex barrier for neural network certification. In: Proc. of
  NeurIPS'19. pp. 15072--15083 (2019)

\bibitem{TranYLMNXBJ20}
Tran, H., Yang, X., Lopez, D.M., Musau, P., Nguyen, L.V., Xiang, W., Bak, S.,
  Johnson, T.T.: {NNV:} the neural network verification tool for deep neural
  networks and learning-enabled cyber-physical systems. In: Proc. of CAV'20.
  LNCS, vol. 12224, pp. 3--17. Springer (2020)

\bibitem{WangPWYJ18}
Wang, S., Pei, K., Whitehouse, J., Yang, J., Jana, S.: Formal security analysis
  of neural networks using symbolic intervals. In: Proc. of {USENIX} Security
  (USENIX). pp. 1599--1614 (2018)

\bibitem{weng2018towards}
Weng, T.W., Zhang, H., Chen, H., Song, Z., Hsieh, C.J., Daniel, L., Dhillon,
  I.: Towards fast computation of certified robustness for relu networks. In:
  International Conference on Machine Learning (ICML) (2018)

\bibitem{zhang2018efficient}
Zhang, H., Weng, T.W., Chen, P.Y., Hsieh, C.J., Daniel, L.: Efficient neural
  network robustness certification with general activation functions. In:
  Proceedings of the 32nd International Conference on Neural Information
  Processing Systems. pp. 4944--4953 (2018)

\end{thebibliography}
\bibliographystyle{splncs03}

\appendix
\clearpage



\section{Additional Experimental Results}

Here, we present additional plots of reachable sets computed by different techniques for the benchmarks in Section 4 of the main paper.

For each benchmark, the goal is to check whether the system will reach a given target set. For each tool and in each test, if the computed reachable set overapproximation for the last control step lies entirely in the target set, we consider the tool to have successfully verified the reachability property. If the overapproximation of the reachable set does not intersect with the target set, the tool would have successfully disproved the reachability property. Otherwise, we consider the verification result to be unknown.

\begin{figure}[tbp]
\captionsetup[subfloat]{farskip=2pt,captionskip=1pt}
	\centering	
	\subfloat[][ex1-relu]{%
		\includegraphics[width=0.23\textwidth]{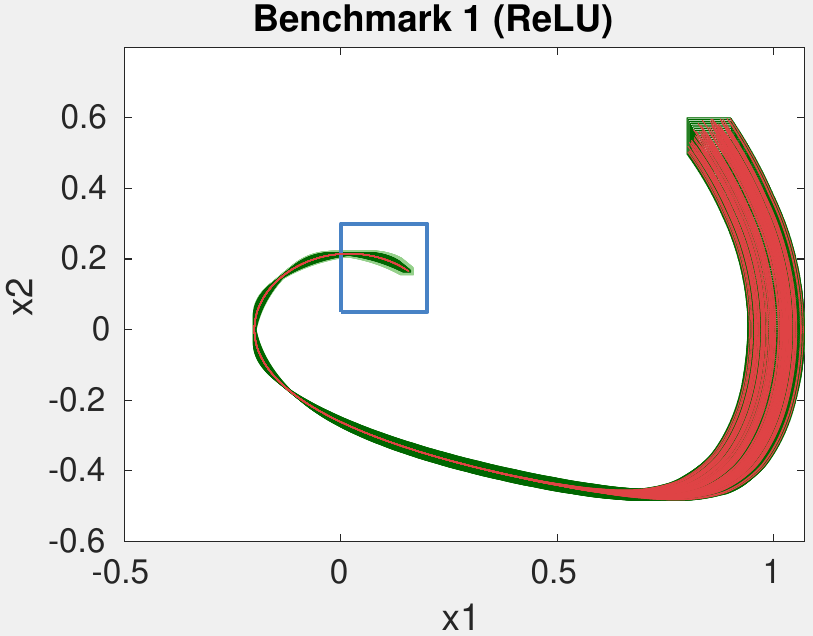}%
		\label{fig:ex1-relu}%
	}\ 
	\subfloat[][ex1-sigmoid]{%
		\includegraphics[width=0.23\textwidth]{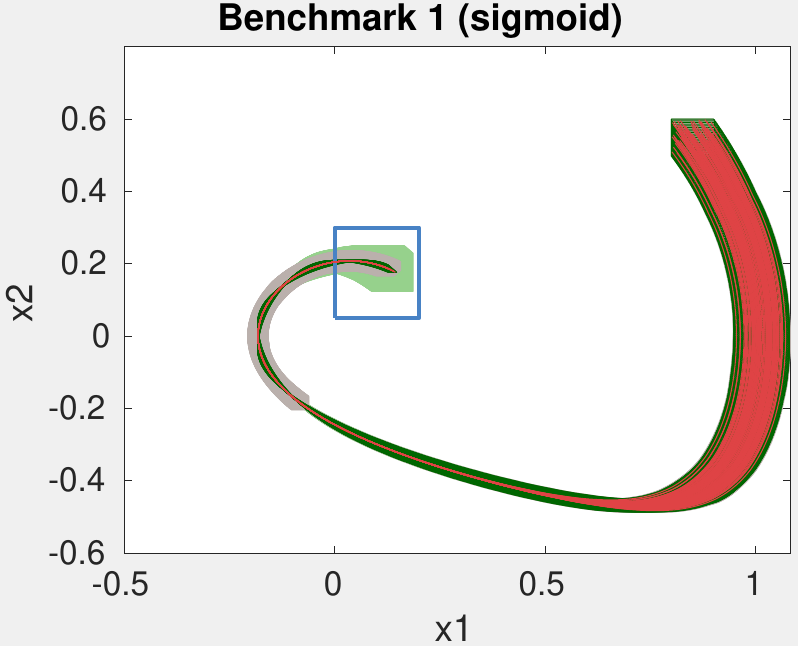}%
		\label{fig:ex1-sigmoid}%
	}\ 
	\subfloat[][ex1-tanh]{%
		\includegraphics[width=0.23\textwidth]{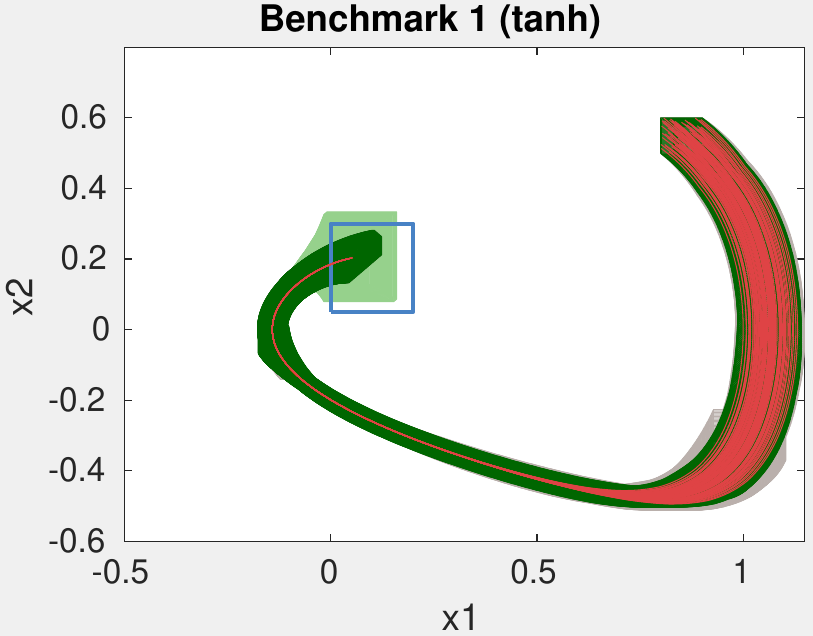}%
		\label{fig:ex1-tanh}%
	}\ 
	\subfloat[][ex1-relu-tanh]{%
		\includegraphics[width=0.23\textwidth]{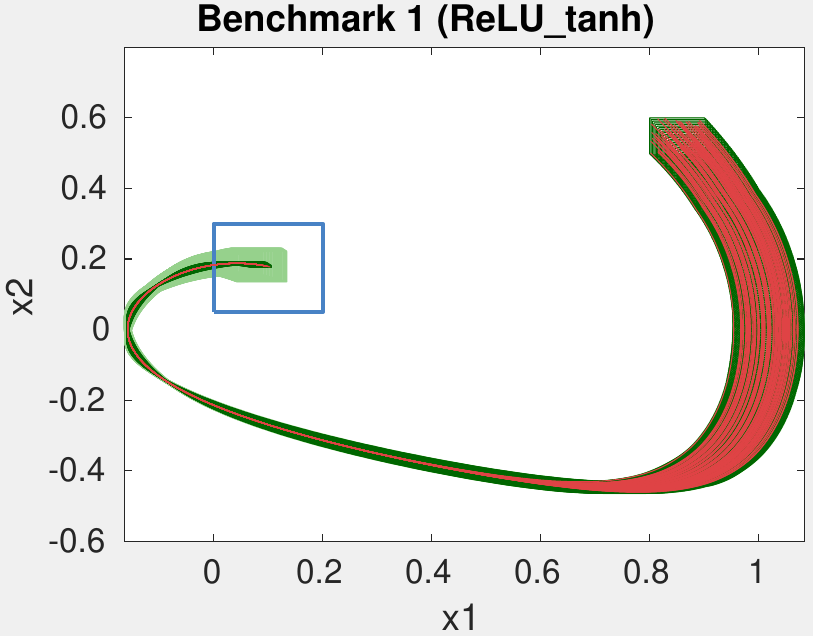}%
		\label{fig:ex1-relu-tanh}%
	}\\
	\subfloat[][ex2-relu]{%
		\includegraphics[width=0.23\textwidth]{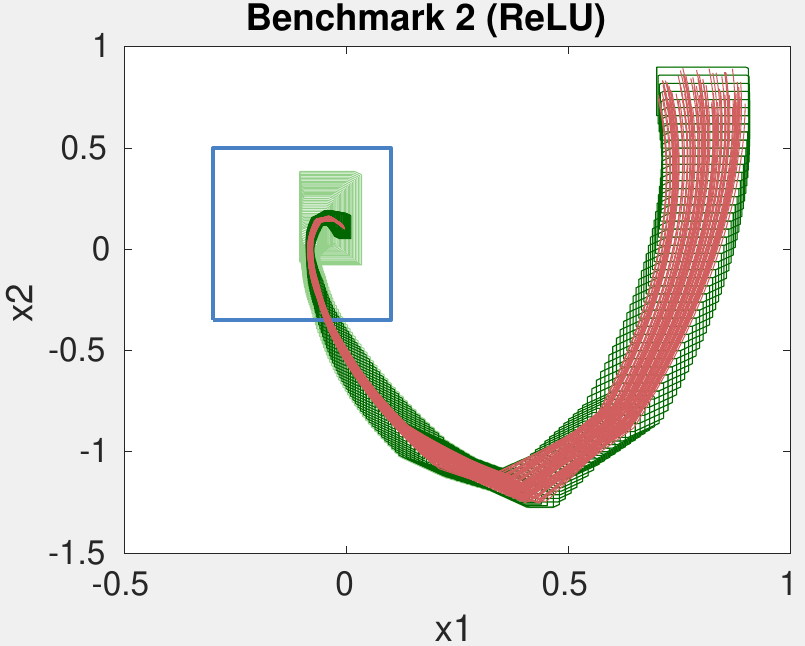}%
		\label{fig:ex2-relu}%
	}\ 
	\subfloat[][ex2-sigmoid]{%
		\includegraphics[width=0.23\textwidth]{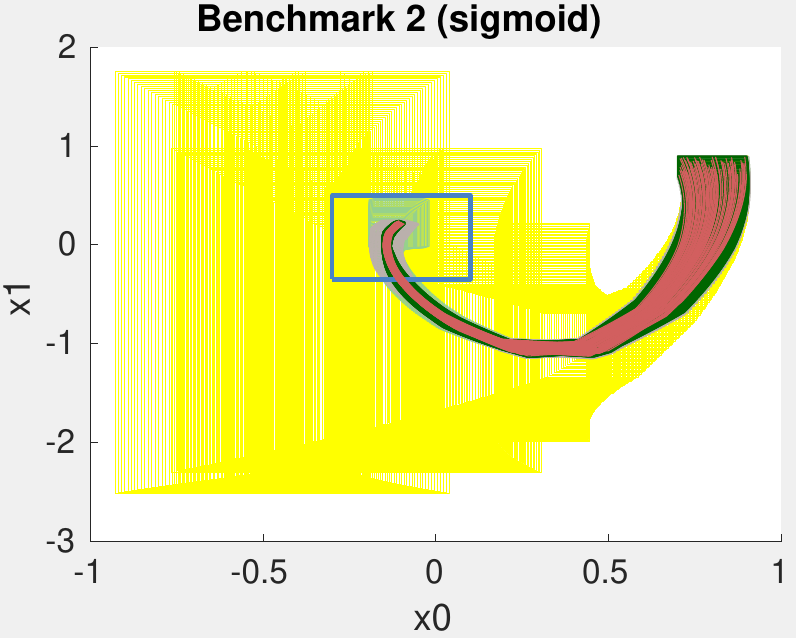}%
		\label{fig:ex2-sigmoid}%
	}\ 
	\subfloat[][ex2-tanh]{%
		\includegraphics[width=0.23\textwidth]{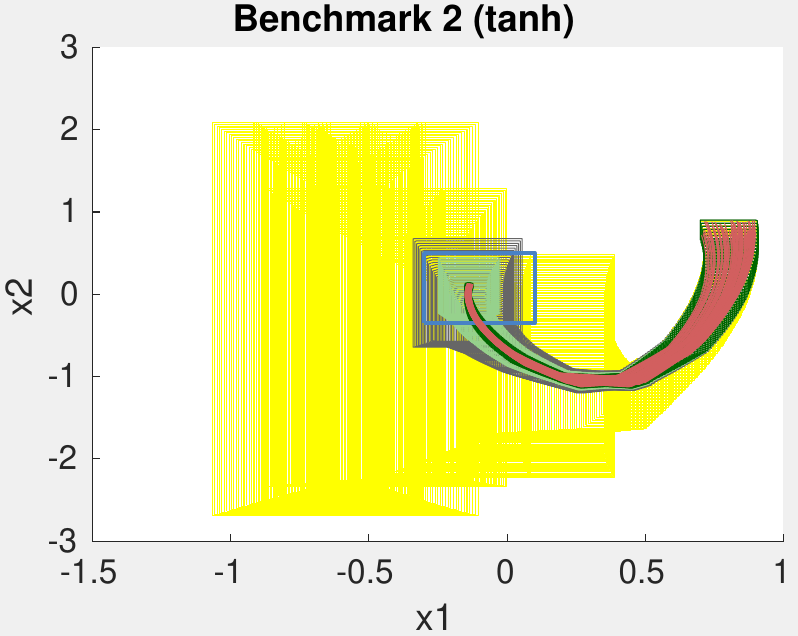}%
		\label{fig:ex2-tanh}%
	}\ 
	\subfloat[][ex2-relu-tanh]{%
		\includegraphics[width=0.23\textwidth]{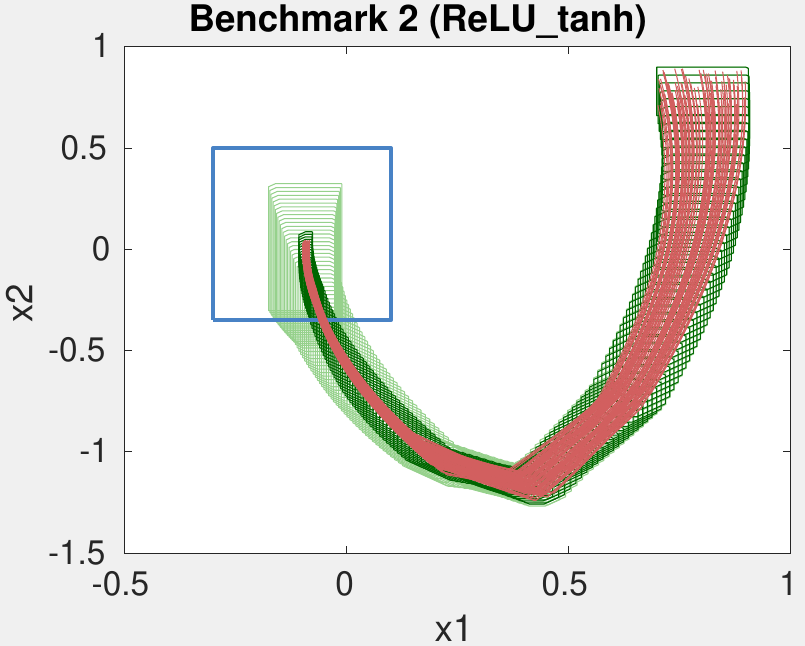}%
		\label{fig:ex2-relu-tanh}%
	}\\
	\subfloat[][ex3-relu]{%
		\includegraphics[width=0.23\textwidth]{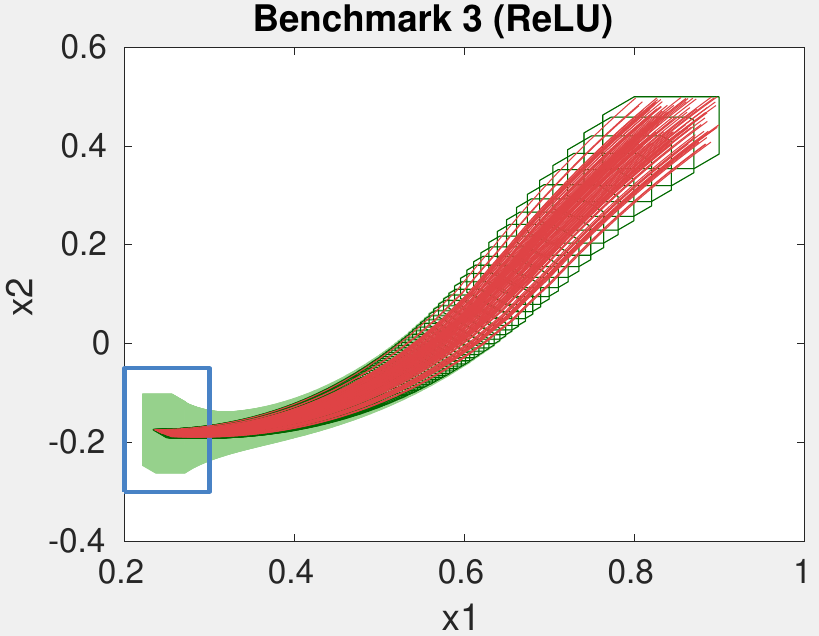}%
		\label{fig:ex3-relu}%
	}\ 
	\subfloat[][ex3-sigmoid]{%
		\includegraphics[width=0.23\textwidth]{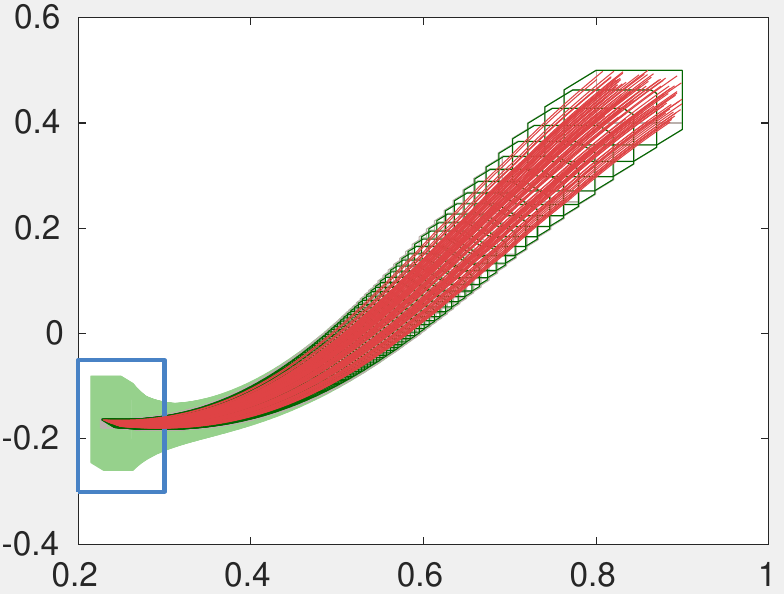}%
		\label{fig:ex3-sigmoid}%
	}\ 
	\subfloat[][ex3-tanh]{%
		\includegraphics[width=0.23\textwidth]{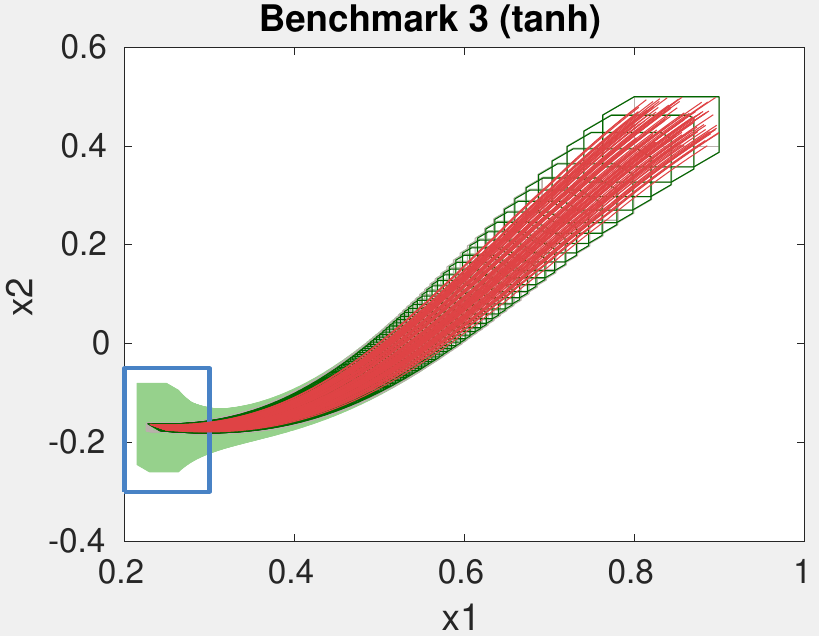}%
		\label{fig:ex3-tanh}%
	}\ 
	\subfloat[][ex3-relu-sigmoid]{%
		\includegraphics[width=0.23\textwidth]{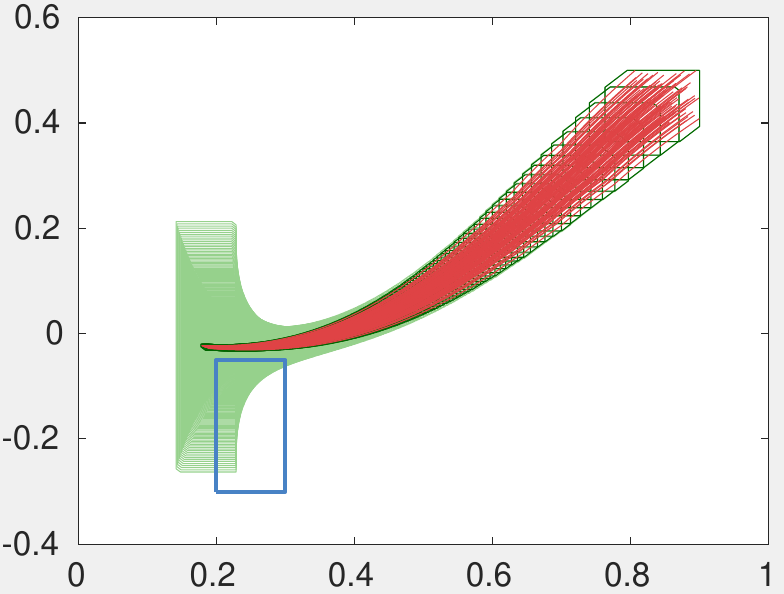}%
		\label{fig:ex3-relu-sigmoid}%
	}\\
	\subfloat[][ex4-relu]{%
		\includegraphics[width=0.23\textwidth]{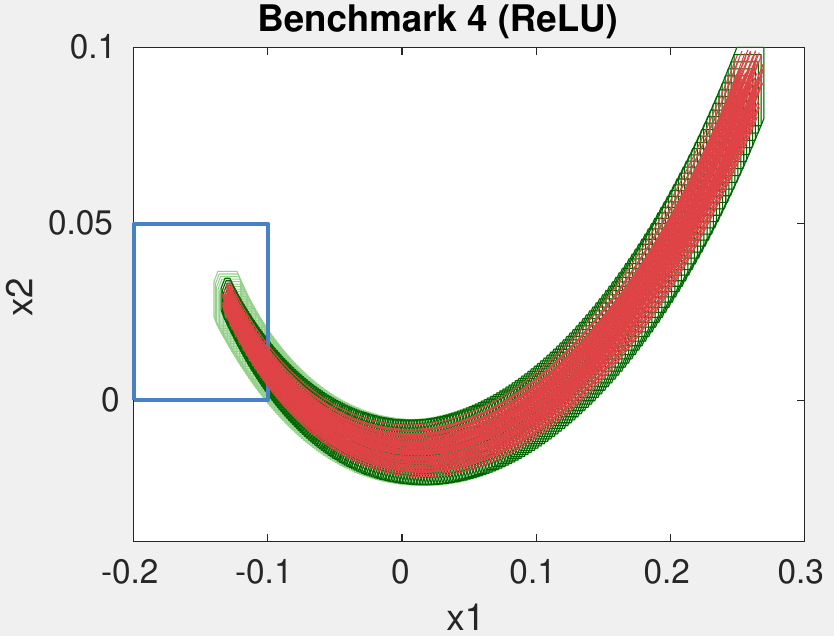}%
		\label{fig:ex4-relu}%
	}\ 
	\subfloat[][ex4-sigmoid]{%
		\includegraphics[width=0.23\textwidth]{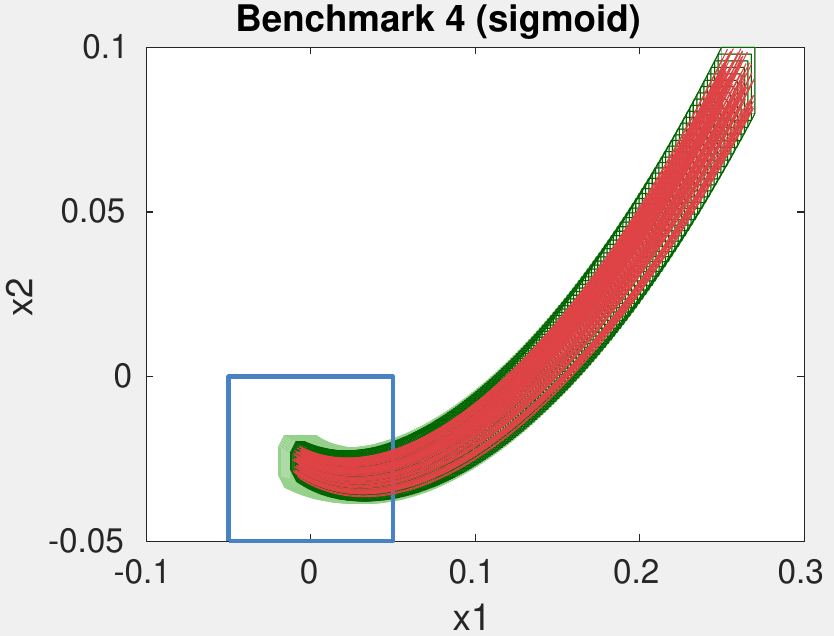}%
		\label{fig:ex4-sigmoid}%
	}\ 
	\subfloat[][ex4-tanh]{%
		\includegraphics[width=0.23\textwidth]{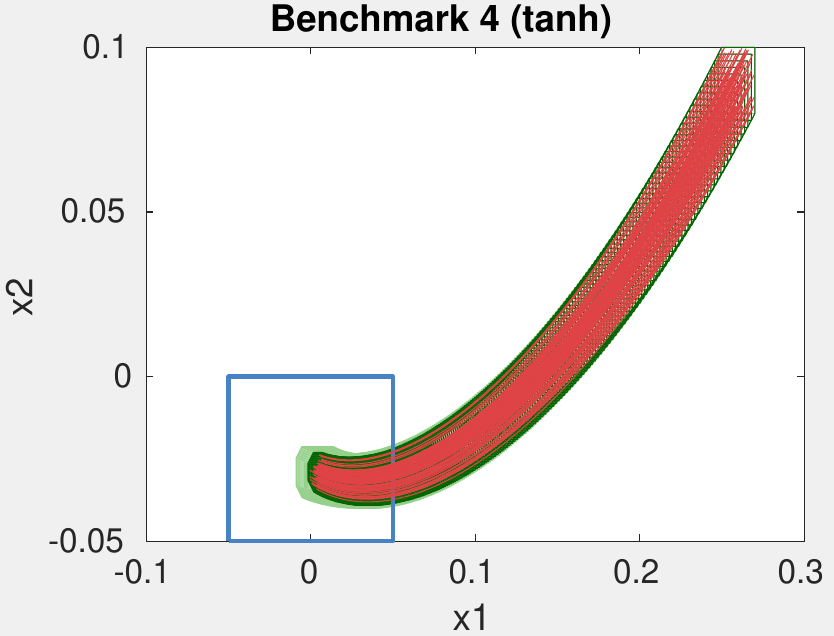}%
		\label{fig:ex4-tanh}%
	}\ 
	\subfloat[][ex4-relu-tanh]{%
		\includegraphics[width=0.23\textwidth]{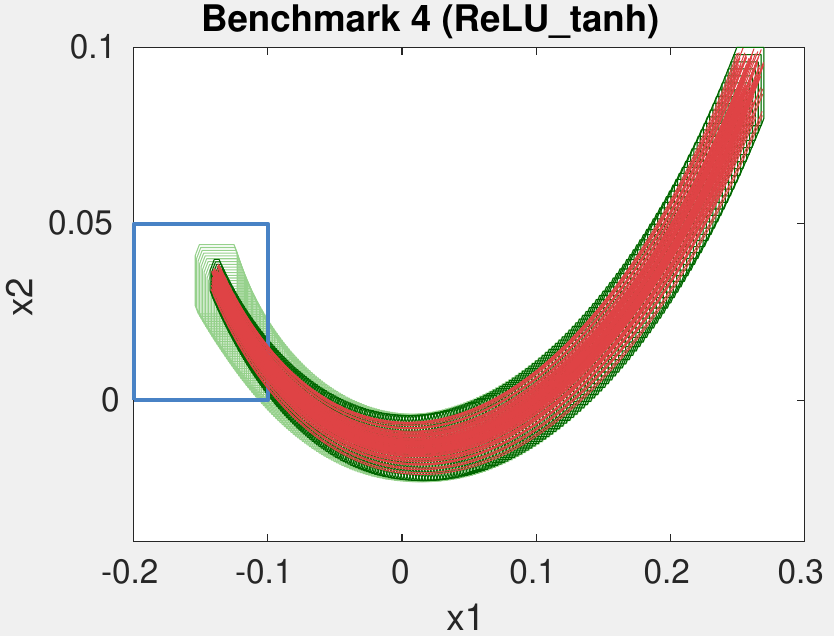}%
		\label{fig:ex4-relu-tanh}%
	}\\
	\subfloat[][ex5-relu]{%
		\includegraphics[width=0.23\textwidth]{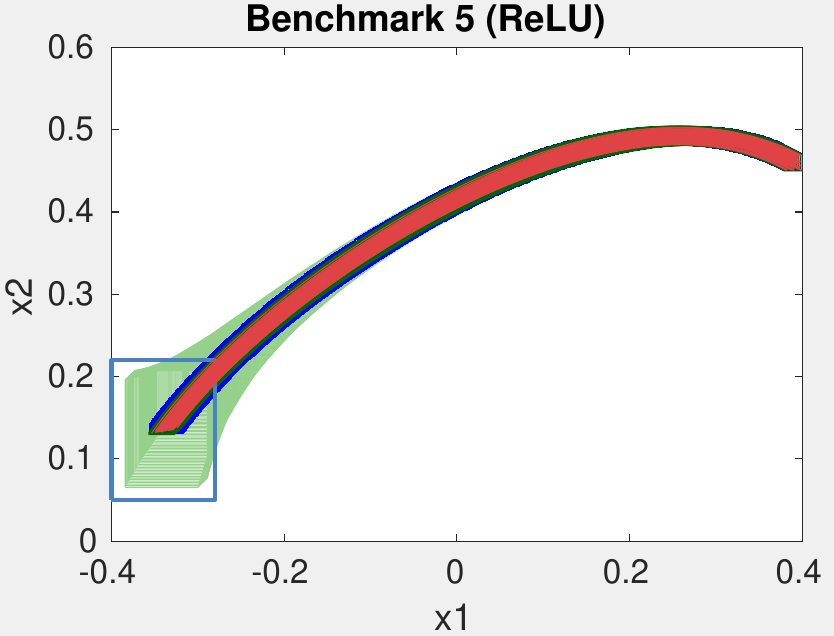}%
		\label{fig:ex5-relu}%
	}\ 
	\subfloat[][ex5-sigmoid]{%
		\includegraphics[width=0.23\textwidth]{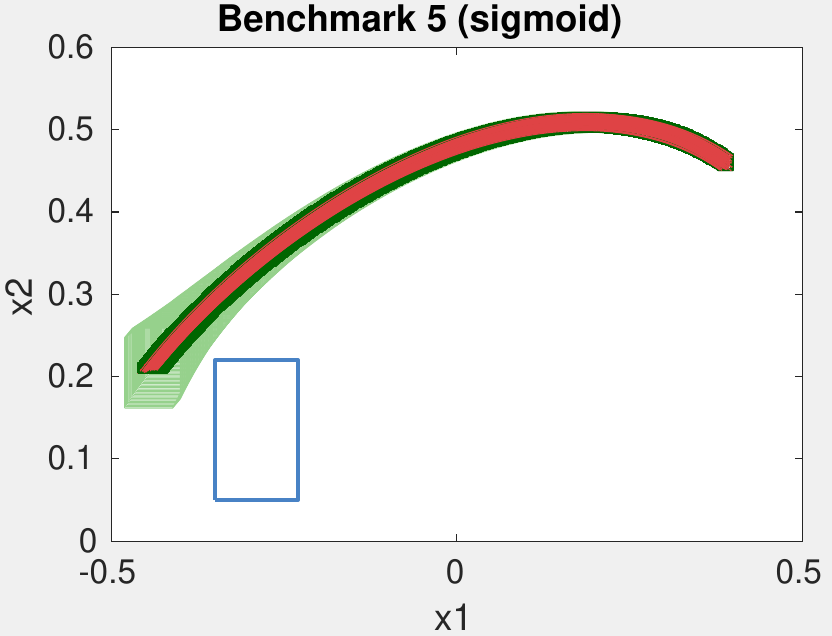}%
		\label{fig:ex5-sigmoid}%
	}\ 
	\subfloat[][ex5-tanh]{%
		\includegraphics[width=0.23\textwidth]{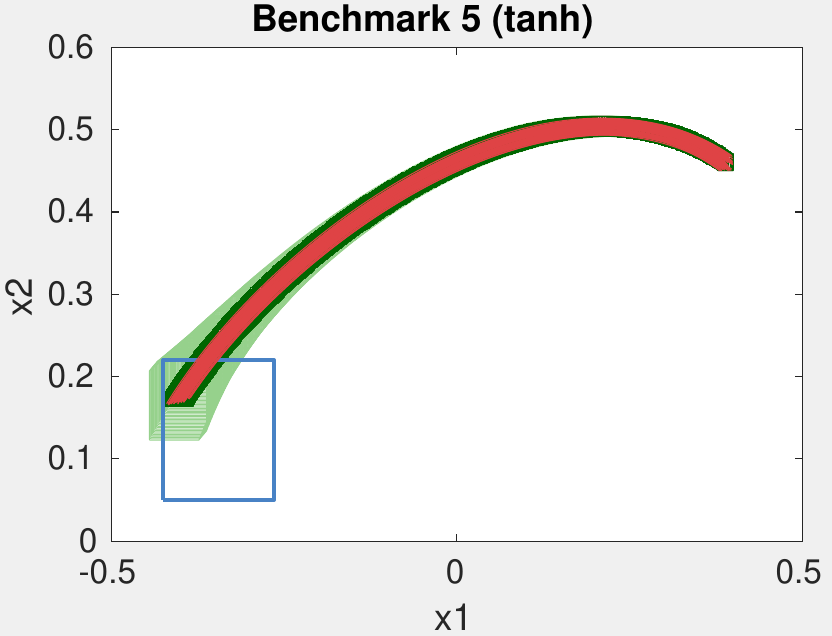}%
		\label{fig:ex5-tanh}%
	}\ 
	\subfloat[][ex5-relu-tanh]{%
		\includegraphics[width=0.23\textwidth]{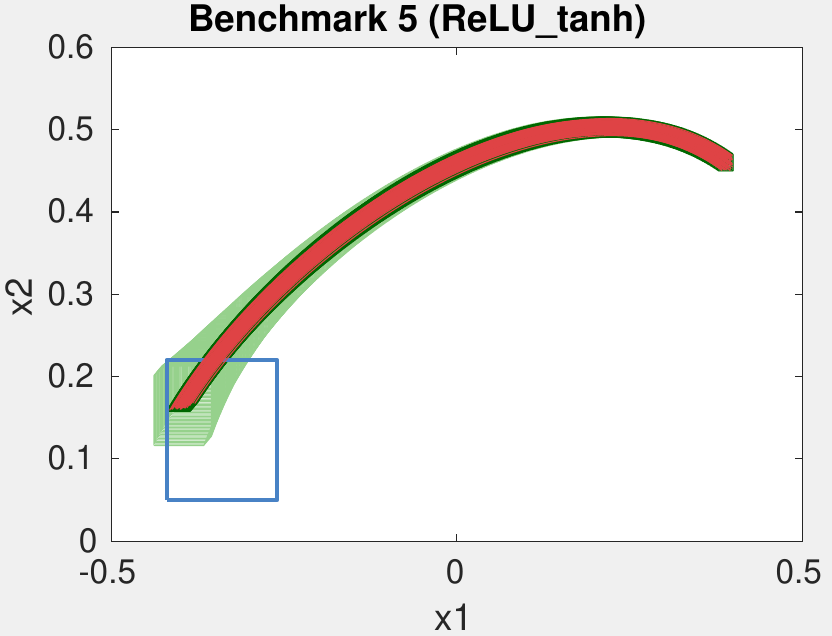}%
		\label{fig:ex5-relu-tanh}%
	}\\
	\subfloat[][ex6-relu]{%
		\includegraphics[width=0.23\textwidth]{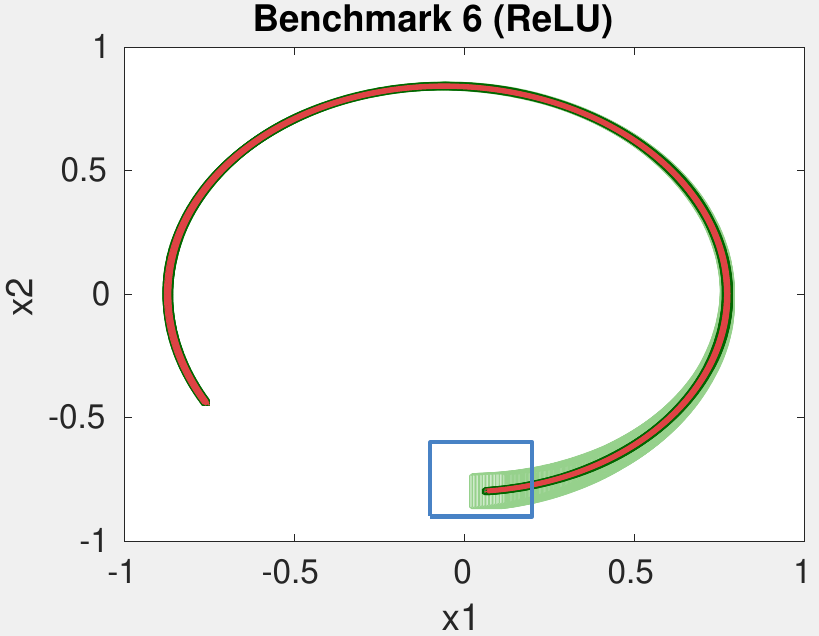}%
		\label{fig:ex6-relu}%
	}\ 
	\subfloat[][ex6-sigmoid]{%
		\includegraphics[width=0.23\textwidth]{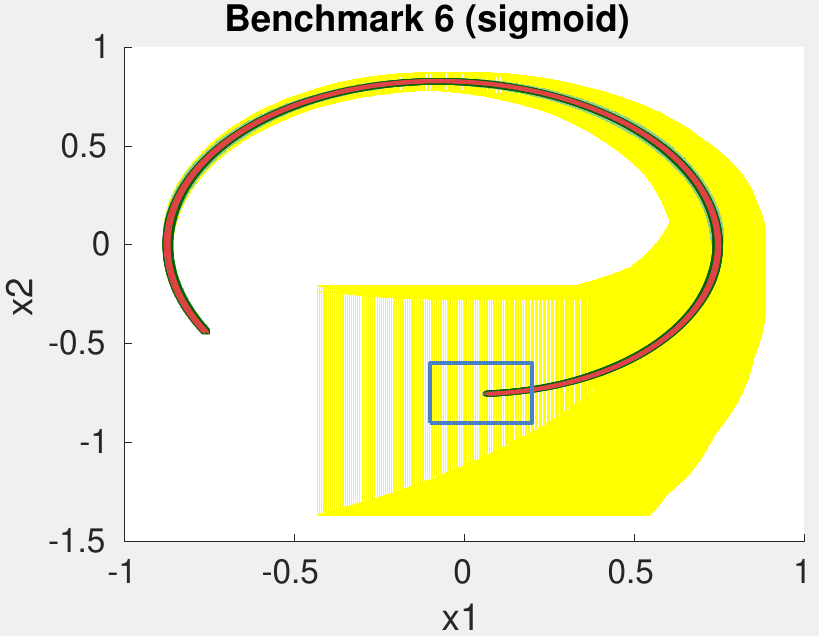}%
		\label{fig:ex6-sigmoid}%
	}\ 
	\subfloat[][ex6-tanh]{%
		\includegraphics[width=0.23\textwidth]{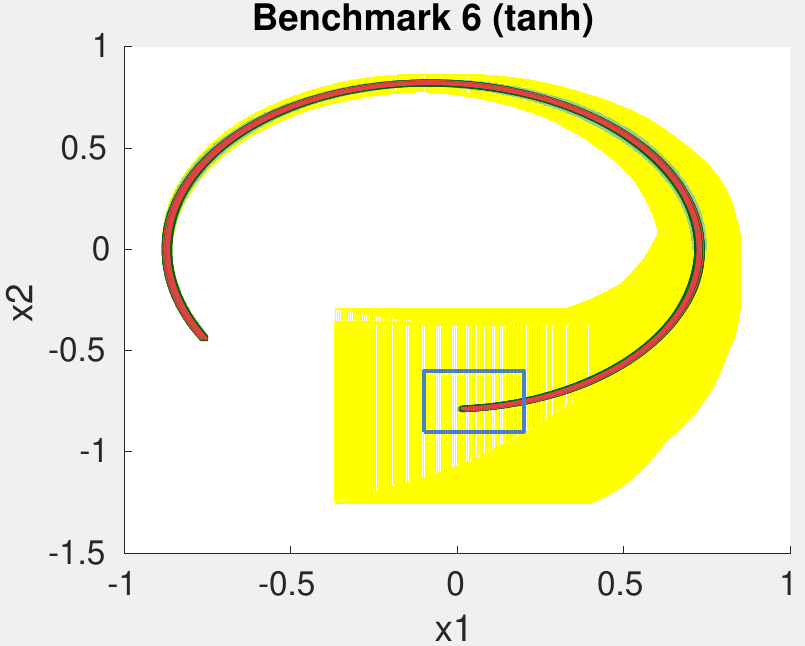}%
		\label{fig:ex6-tanh}%
	}\ 
	\subfloat[][ex6-relu-tanh]{%
		\includegraphics[width=0.23\textwidth]{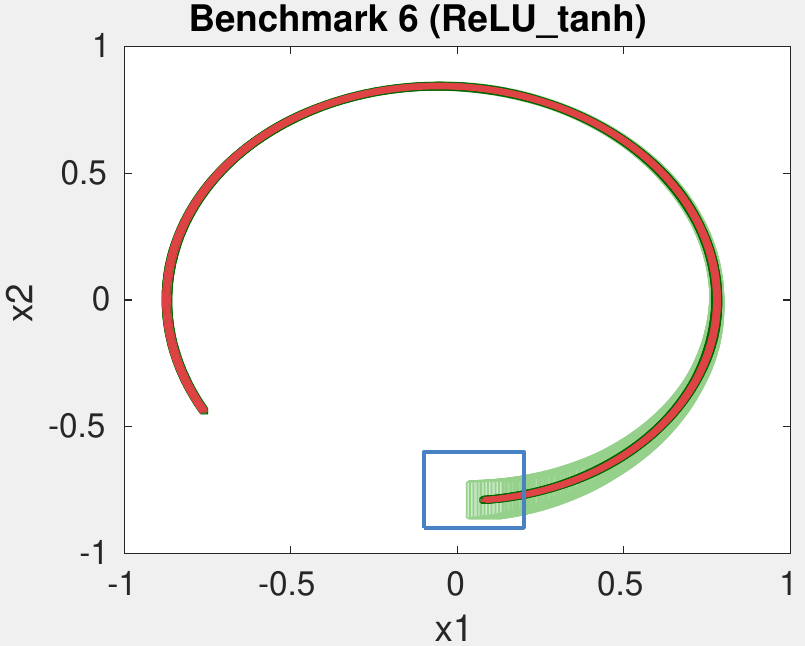}%
		\label{fig:ex6-relu-tanh}%
	}
	\caption{Results of Benchmarks. Except for (f), POLAR produces the tightest reachable set estimation (dark green sets) and successfully proves or disproves the reachability property for all the examples. This is in comparison with other STOA tools including  ReachNN*~\cite{HuangFLC019,FanHCL020} (light green sets), Sherlock~\cite{DuttaCS19} (blue sets), Verisig 2.0~\cite{IvanovCWAPL21} (grey sets), and NNV~\cite{TranYLMNXBJ20} (yellow sets). Except for (f), (g), (v) and (w), NNV used up the memory and couldn't finish the computation.}
	\label{fig:appendix_simulation}
\end{figure}

\begin{figure}[ht!]
    \centering
    \includegraphics[width=0.5\columnwidth]{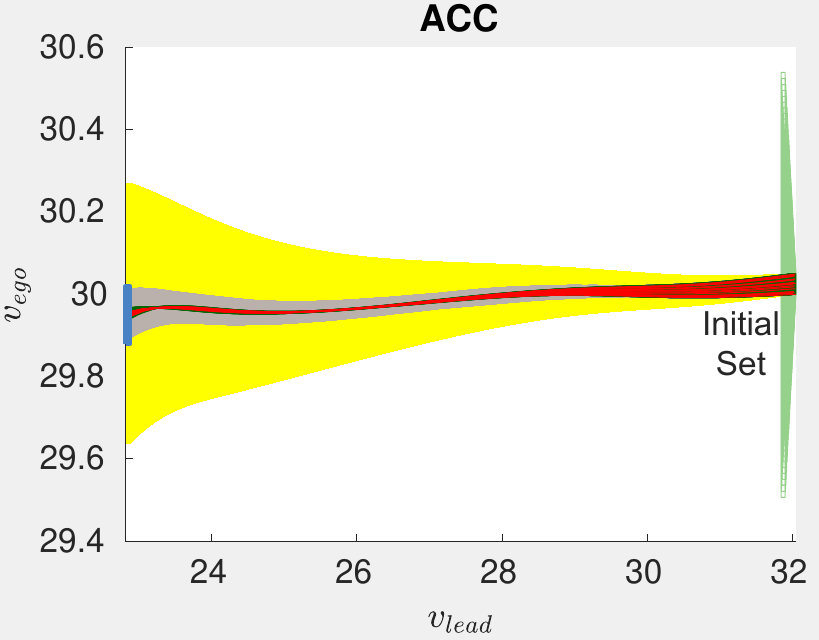}
    \caption{Results of Adaptive Cruise Control (ACC). POLAR for 50 steps (dark green sets), Verisig 2.0 for 50 steps (grey sets), ReachNN* for 3 steps (light green sets), NNV for 50 steps (yellow sets), and simulation traces for 50 steps (red curves).}
    \label{fig:acc}
\end{figure}

\begin{figure}[ht!]
\captionsetup[subfloat]{farskip=2pt,captionskip=1pt}
	\centering	
	\subfloat[][]{%
		\includegraphics[width=0.49\textwidth]{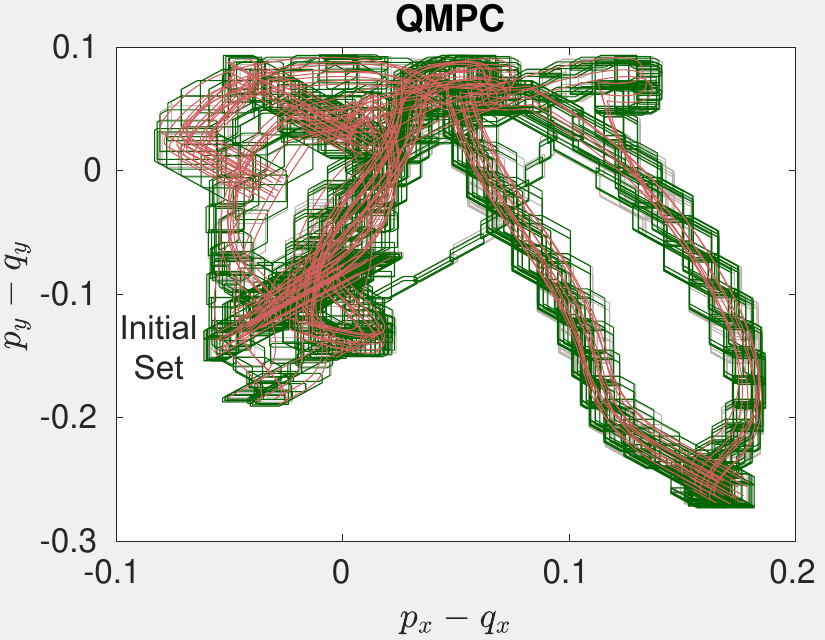}%
		\label{fig:quadrotor_1}%
	}\ 
	\subfloat[][]{%
		\includegraphics[width=0.49\textwidth]{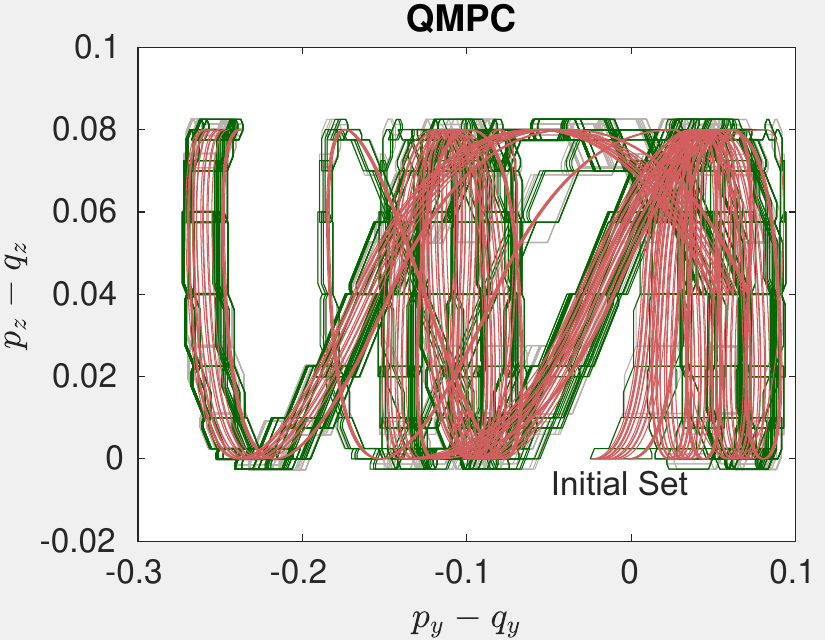}%
		\label{fig:quadrotor_2}%
	}
	\caption{Results of QMPC. POLAR for 30 steps (dark green sets), Verisig 2.0 for 30 steps (grey sets), and simulation traces for 30 steps (red curves).}
	\label{fig:qmpc}
\end{figure}

Results of Benchmark 1-6, the ACC benchmark, and the QMPC benchmark are shown in Figure \ref{fig:appendix_simulation}, \ref{fig:acc}, \ref{fig:qmpc} respectively, while the results of the Attitude control benchmark and the QUAD benchmark are shown previously in Figure~\ref{fig:attitude-control} and Figure \ref{subfig:quad}. The red trajectories are sample system executions and should be contained entirely by the flowpipes computed by each tool.
The dark green sets are the flowpipes computed by POLAR. The light green sets are the flowpipes computed by ReachNN*~\cite{HuangFLC019,FanHCL020}. The blue sets are the flowpipes computed by Sherlock~\cite{DuttaCS19}. The grey sets are the flowpipes computed by Verisig 2.0~\cite{IvanovCWAPL21}. In some benchmarks, the reachable sets computed by Verisig 2.0 are almost overlapping with the reachable sets computed by POLAR. However, POLAR takes much less time to compute the reachable sets compared to Verisig 2.0 as shown in Table 1 of the main paper. We also show results from NNV~\cite{TranYLMNXBJ20} in yellow for some of the benchmarks. For the rest, NNV used up all of the system memory (8GB) and could not finish the computation. Our observations are consistent with those in \cite{IvanovCWAPL21} where NNV is not able to verify any of these benchmarks.
The blue box represents the target set in each test. POLAR produces the tightest reachable set estimation and successfully proves or disproves the reachability property for all the examples.


\newpage

\section{Discrete-time Mountain Car Benchmark} \label{app_sec:mc}
\begin{example}[MC] \label{ex:mountain}
In this benchmark, an under-powered car targets to drive up a steep hill. Since the car does not have enough power to Widthelerate up the hill, it needs to drive up the opposite hill first to gain enough momentum. The car has the following discrete-time dynamics:
\begin{displaymath}
\left\{
\begin{aligned}
    x_{0}[t+1] &= x_{0}[t] + x_{1}[t], \\
    x_{1}[t+1] &= x_{1}[t] + 0.0015 \cdot u[t] - 0.0025 \cdot \cos(3 \cdot x_{0}[t]).
\end{aligned}
\right.
\end{displaymath}
For this benchmark, the initial set is $x_0 \in [-0.53, -0.5]$ and $x_1 = 0$. The target is $x_0 \ge 0.2$ and $x_1 \ge 0$ where the car reaches the top of the hill and is moving forward. The total control steps $N$ is 150.
\end{example}

\section{Benchmarks with High Dimensional States and Multiple Outputs} \label{app_sec:ac_quad}
\begin{example}[Attitude Control] \label{ex:attitude}
We consider the attitude control of a rigid body with six states and three inputs as a physically illustrating example \cite{prajna2004nonlinear}. The system dynamics is
\begin{displaymath}
\left\{
\begin{aligned}
&\dot{\omega}_1 = 0.25({u_0} + {\omega_2\omega_3}),  \qquad
\dot{\omega}_2 = 0.5({u_1} - {3\omega_1\omega_3}), \qquad
\dot{\omega}_3 = u_2 + 2\omega_1\omega_2, \\
&\dot{\psi}_1 {=} 0.5\left(\omega_2({\psi_1^2} {+} {\psi_2^2} {+} {\psi_3^2} {-} {\psi_3}) {+} \omega_3({\psi_1^2} {+} {\psi_2^2} {+} {\psi_2} {+} {\psi_3^2}) {+} \omega_1({\psi_1^2} {+} {\psi_2^2} {+} {\psi_3^2} {+} {1})\right), \\
&\dot{\psi}_2 {=} 0.5\left(\omega_1({\psi_1^2} {+} {\psi_2^2} {+} {\psi_3^2} {+} {\psi_3}) {+} \omega_3({\psi_1^2} {-} {\psi_1} {+} {\psi_2^2} {+} {\psi_3^2}) {+} \omega_2({\psi_1^2} {+} {\psi_2^2} {+} {\psi_3^2} {+} {1})\right), \\
&\dot{\psi}_3 {=} 0.5\left(\omega_1({\psi_1^2} {+} {\psi_2^2} {-} {\psi_2} {+} {\psi_3^2}) {+} \omega_2({\psi_1^2} {+} {\psi_1} {+} {\psi_2^2} {+} {\psi_3^2}) {+} \omega_3({\psi_1^2} {+} {\psi_2^2} {+} {\psi_3^2} {+} {1})\right).    
\end{aligned}
\right.
\end{displaymath}
wherein the state $\vx {=} (\omega,\psi)$ consists of the angular velocity vector in a body-fixed frame $\omega {\in} \reals^3$, and the Rodrigues parameter vector $\psi {\in} \reals^3$.

The control torque $u {\in} \reals^3$ is updated every $0.1$ second by a neural network with 3 hidden layers, each of which has 64 neurons. The activations of the hidden layers are sigmoid and identity,  respectively. We train the neural-network controller using supervised learning methods to learn from a known nonlinear controller~\cite{prajna2004nonlinear}.
The initial state set is:
\begin{align*}
&\omega_1\in [-0.45, -0.44], \omega_2\in [-0.55, -0.54], \omega_3\in [0.65, 0.66], \\
&\psi_1\in [-0.75, -0.74], \psi_2 \in [0.85, 0.86], \psi_3 \in [-0.65, -0.64].
\end{align*}
\end{example}

\begin{example}[QUAD]
We study a neural-network controlled quadrotor (QUAD) with 12 states~\cite{beard2008quadrotor}. For the states, we have the inertial (north) position $x_1$, the inertial (east) position $x_2$, the altitude $x_3$, the longitudinal velocity $x_4$, the lateral velocity $x_5$, the vertical velocity $x_6$, the roll angle $x_7$, the pitch angle $x_8$, the yaw angle $x_9$, the roll rate $x_{10}$, the pitch rate $x_{11}$, and the yaw rate $x_{12}$. The control torque $u \in \reals^3$ is updated every 0.1 second by a neural network with 3 hidden layers, each of which has 64 neurons. The activations of the hidden layers and the output layer are sigmoid and identity, respectively.

\begin{displaymath}
\left\{
\begin{aligned}
\dot{x}_1 = & \cos(x_8) \cos(x_9) x_4 + \left ( \sin(x_7) \sin(x_8) \cos(x_9) - \cos(x_7) \sin(x_9) \right ) x_5 \\ & + \left ( \cos(x_7) \sin(x_8) \cos(x_9) + \sin(x_7) \sin(x_9) \right) x_6 \\
\dot{x}_2 = & \cos(x_8) \sin(x_9) x_4 + \left ( \sin(x_7) \sin(x_8) \sin(x_9) + \cos(x_7) \cos(x_9) \right) x_5 \\ & + \left ( \cos(x_7) \sin(x_8) \sin(x_9) - \sin(x_7) \cos(x_9) \right) x_6 \\
\dot{x}_3 = & \sin(x_8) x_4 - \sin(x_7) \cos(x_8) x_5 - \cos(x_7) \cos(x_8) x_6 \\
\dot{x}_4 = & x_{12} x_5 - x_{11} x_6 - g \sin(x_8) \\
\dot{x}_5 =  & x_{10} x_6 - x_{12} x_4 + g \cos(x_8) \sin(x_7) \\
\dot{x}_6 = & x_{11} x_4 - x_{10} x_5 + g \cos(x_8) \cos(x_7) - g - u_1 / m \\
\dot{x}_7 = & x_{10} + \sin(x_7) \tan(x_8) x_{11} + \cos(x_7) \tan(x_8) x_{12} \\
\dot{x}_8 = & \cos(x_7) x_{11} - \sin(x_7) x_{12} \\
\dot{x}_{9} = & \frac{\sin(x_7)}{\cos(x_8)} x_{11} - \sin(x_7) x_{12} \\
\dot{x}_{10} = & \frac{J_y - J_z}{J_x} x_{11} x_{12} + \frac{1}{J_x} u_2 \\
\dot{x}_{11} = & \frac{J_z - J_x}{J_y} x_{10} x_{12} + \frac{1}{J_y} u_3 \\
\dot{x}_{12} = & \frac{J_x - J_y}{J_z} x_{10} x_{11} + \frac{1}{J_z} \tau_\psi
\end{aligned}
\right.
\end{displaymath}

The initial set is:
\begin{align*}
&x_1 {\in} [-0.4, 0.4], x_2 {\in} [-0.4, 0.4], x_3 {\in} [-0.4, 0.4], x_4 {\in} [-0.4, 0.4], \\ & x_5 {\in} [-0.4, 0.4], x_6 {\in} [-0.4, 0.4], x_7 {=} 0, x_8 {=} 0, x_9 {=} 0, x_{10} {=} 0, x_{11} {=} 0, x_{12} {=} 0
\end{align*}

The control goal is to stabilize the attitude $x_3$ to a goal region $[0.94, 1.06]$. 
\end{example}

\begin{example}[Discrete-Time Mountain Car (MC)]
We consider a common benchmark in Reinforcement Learning problems, namely Mountain Car. In this benchmark, an under-powered car targets to drive up a steep hill. Since the car does not have enough power to accelerate up the hill, it needs to drive up the opposite hill first to gain enough momentum. The car has the following discrete-time dynamics:
\begin{align*}
    x_{0}[t+1] &= x_{0}[t] + x_{1}[t], \\
    x_{1}[t+1] &= x_{1}[t] + 0.0015 \cdot u[t] - 0.0025 \cdot \cos(3 \cdot x_{0}[t]).
\end{align*}

For this benchmark, the initial set is $x_0 \in [-0.53, -0.5]$ and $x_1 = 0$. The target is $x_0 \ge 0.2$ and $x_1 \ge 0$ where the car reaches the top of the hill and is moving forward. The total control steps $N$ is 150.
\end{example}

\section{Theorem Proof}

\subsection{Proof of Soundness of Sampling-based Error Analysis}
\textbf{Proof.} The input range of an activation function $\sigma_j$ is subdivided into $m$ line segments. Consider the $i$-th segment $[\frac{\overline{Z}_j-\underline{Z}_j}{m}(i-1)+\underline{Z}_j,\frac{\overline{Z}_j-\underline{Z}_j}{m}(i)+\underline{Z}_j]$, and let $c=\frac{\overline{Z}_j-\underline{Z}_j}{m}(i-\frac{1}{2})+\underline{Z}_j$ be the center of the segment. The difference between the Bernstein polynomial $p_{\sigma}^j$ and the activation function at the center of the $i$-th segment is computed as
$
 \left|p_{\sigma}^j(c) - \sigma_j(c)\right|.
$
Then, the value of $\epsilon_j$ can be bounded by this difference at the center, as well as the product between the Lipschitz constant of the activation function with respect to this segment $L_j$ and the size of the segment $\frac{\overline{Z}_j-\underline{Z}_j}{m}$, i.e., $L_j\cdot\frac{\overline{Z}_j-\underline{Z}_j}{m}$. The detailed deduction is given below.
$$
 \small
 \begin{aligned}
 & |p_{\sigma_j,i}(x)-\sigma_j(x)| & \text{}\\
 = & | p_{\sigma_j,i}(x) - p_{\sigma+j,i}(c) + p_{\sigma_j,i}(c)-\sigma_j(c) + \sigma_j(c) - \sigma_j(x) | & \text{}\\
 \leq & |p_{\sigma_j,i}(x) {-} p_{\sigma_j,i}(c)| {+} |p_{\sigma_j,i}(c){-}\sigma_j(c)| {+} |\sigma_j(c) {-} \sigma_j(x)| & \text{Triangle inequality}\\
 \leq & |p_{\sigma_j,i}(x) {-} p_{\sigma_j,i}(c)| {+} |p_{\sigma_j,i}(c){-}\sigma_j(c)| {+} L_j\cdot\frac{\overline{Z}_j{-}\underline{Z}_j}{2m} & \text{Lipschitz continuity for } \sigma_j\\
 \leq & L_j\cdot\frac{\overline{Z}_j{-}\underline{Z}_j}{2m} {+} |p_{\sigma_j,i}(c){-}\sigma_j(c)| {+} L_j\cdot\frac{\overline{Z}_j{-}\underline{Z}_j}{2m} & \text{Lipschitz continuity for } p_{\sigma_j,i}\\
 = & |p_{\sigma_j,i}(c)-\sigma_j(c)| + L_j\frac{\overline{Z}_j-\underline{Z}_j}{m} & \text{}
 \end{aligned}
$$
Note that Bernstein polynomial $p_{\sigma_j,i}$ has the same Lipschitz constant with $\sigma_j$. Thus we also use $L_j$ to bound $|p_{\sigma_j,i}(x) - p_{\sigma_j,i}(c)|$ in the deduction. The error bound over the whole range $[\underline{Z}_j,\overline{Z}_j]$ should be the largest error bound among all the segments. \hfill $\square$.

\subsection{Proof of Theorem 1}
\textbf{Proof.} First, due to the overapproximation property of our methods, any of our Bernstein overapproximation $p_{\sigma,I} + I_{\sigma,i}$ satisfied that for $\vz$ in the domain on which $I_{\sigma,i}$ is evaluated, we have that $\sigma(\vz) \in p_{\sigma,i} (\vz)  + I_{\sigma,i}$. Therefore, by the overapproximation property of TM arithmetic, the returned $(p_r(\vx_0),I_r)$ of Algorithm~\ref{algo:nn_output} or~\ref{algo:sym_rem} is a state-wise overapproximation of the control input range w.r.t. the TM variable $\vx_0 \in X_0$ wherein $X_0$ is the NNCS initial set.

We prove Theorem 1 by an induction on the number of control steps $j$. Assume that $N = \delta_c / \delta$ is the number of flowpipes computed in each control step.

\noindent\textbf{Base Case.} When $j=1$, the TM flowpipes are computed for the reachable set in the first control step and the evolution is under the pure continuous dynamics $\dot{\vx} = f(\vx,\vvu_0)$, $\dot{\vvu} = 0$ with $\vx(0)\in X_0$ and $\vvu(0) = \kappa(\vx(0))$. The image of the mapping $\kappa(\vx_0)$ from $\vx_0\in X_0$ is overapproximated by a TM $(p_r(\vx_0),I_r)$ with $\vx_0\in X_0$. Hence, by performing TM flowpipe construction for the ODE $\dot{\vx} = f(\vx,\vvu)$, $\dot{\vvu} = 0$ with the initial set $\vvu(0)\in p_r(\vx_0) + I_r$, $\vx(0) = \vx_0$, we have that for any $i=1,\dots,N$, the $i$-th TM flowpipe $\mathcal{F}_i(\vx_0,\tau)$ contains the exact reachable state at the time $(i-1)\delta + \tau$ for $\tau\in [0,\delta]$.

\noindent\textbf{Induction.} When $j>1$, we assume that the local initial set $\hat{X}_{j-1} = (p_0(\vx_0),I_0)$ is a state-wise overapproximation of the reachable set at the time $j\delta_c$ from any $\vx_0\in X_0$. Then the TM $(p_r(\vx_0),I_r)$ is a state-wise overapproximation for the control input set $\kappa(\hat{X}_{j-1})$, i.e., the NN controller's output produced based on the $j\delta_c$-time state in the execution from an initial state $\vx_0\in X_0$ is contained in the box $p_r(\vx_0) + I_r$ for any $\vx_0\in X_0$. Hence, for any $i=1,\dots,N$, the $i$-th flowpipe $\mathcal{F}_i(\vx_0,\tau)$ computed for the ODE $\dot{\vx} = f(\vx,\vvu)$, $\dot{\vvu} = 0$ with the initial set $\vvu(0)\in p_r(\vx_0) + I_r$, $\vx(0) \in \hat{X}_{j-1}$ contains the actual reachable state $\varphi_{\mathcal{N}}(\vx_0,(j-1)\delta_c + (i-1)\delta + \tau)$ for any $\tau\in [0,\delta]$. \hfill$\Box$

\end{document}